\documentclass[asa,12pt]{article}
\usepackage{graphicx}
\usepackage{epstopdf}
\usepackage{amssymb,latexsym,amsmath}     % Standard packages
\usepackage{indentfirst}
\usepackage{booktabs} % Allows the use of \toprule, \midrule and \bottomrule in tables for horizontal lines
\usepackage{pdflscape}
\usepackage{pdfpages}
\usepackage{multicol}
\usepackage{lscape}
\usepackage{tabularx}
\usepackage{bbm}
\usepackage{color, colortbl}

\renewcommand{\theequation}{\arabic{equation}}
\advance\topmargin by -0.75truein
\advance\textheight by 1.25truein
\def\mybibliography#1{{\noindent \Large \bf References}\list
 {}{\setlength{\leftmargin}{1em}\setlength{\labelsep}{0pt}
\itemindent=-\leftmargin}
 \def\newblock{\hskip .02em plus .20em minus -.07em}
 \sloppy\clubpenalty4000\widowpenalty4000
 \sfcode`\.=1000\relax}
\newbox\TempBox \newbox\TempBoxA
\def\uw#1{%
  \ifmmode\setbox\TempBox=\hbox{$#1$}\else\setbox\TempBox=\hbox{#1}\fi%
  \setbox\TempBoxA=\hbox to \wd\TempBox{\hss\char'176\hss}%
  \rlap{\copy\TempBox}\smash{\lower9pt\hbox{\copy\TempBoxA}}%
}
\newbox\TempBox \newbox\TempBoxA
\def\uwd#1{%
  \ifmmode\setbox\TempBox=\hbox{$#1$}\else\setbox\TempBox=\hbox{#1}\fi%
  \setbox\TempBoxA=\hbox to \wd\TempBox{\hss\char'176\hss}%
  \rlap{\copy\TempBox}\smash{\lower10pt\hbox{\copy\TempBoxA}}%
}
\def\mathunderaccent#1{\let\theaccent#1\mathpalette\putaccentunder}
\def\putaccentunder#1#2{\oalign{$#1#2$\crcr\hidewidth
\vbox to.2ex{\hbox{$#1\theaccent{}$}\vss}\hidewidth}}

\begin{document}
\newtheorem{theorem}{Theorem}[section]
\newtheorem{proposition}{Proposition}[section]
\newtheorem{corollary}{Corollary}[section]
\newtheorem{lemma}{Lemma}[section]

\vspace*{.10in}

\vspace*{.10in}

\begin{center} {\large \bf Bayesian Predictive Inference When Integrating\\
 a  Non-probability 
Sample and a Probability Sample}

\vspace*{.25in}

{Balgobin Nandram}\\
\medskip
Department of Mathematical Sciences, Worcester Polytechnic Institute\\
100 Institute Road, Worcester, MA 01609\\
%(balnan@wpi.edu \& Balgobin.Nandram@nass.usda.gov)\\
(balnan@wpi.edu)\\

\vspace*{.25in}

{J. N. K. Rao}\\
\medskip
School of Mathematics and Statistics, Carleton University\\
 Ottawa, Ontario, K1S 5B6, Canada\\
(jrao34@rogers.com)\\

\vspace*{.25in}

%January 24, 2022

%February 12, 2022

%February 19, 2022

%March 12, 2022

%April 8, 2022

April 6, 2023

\end{center}

\begin{center}
{\bf Abstract}
\end{center}

We consider the problem of  integrating  a small probability sample (ps) and a non-probability 
sample (nps). By definition, for the nps, there are no survey weights, but for the ps, there are 
survey weights. The key issue is that the nps, although much larger than the ps, can lead to a 
biased estimator of a finite population quantity but with much smaller variance. We begin with
a relatively simple problem in which the population is assumed to be homogeneous and there are no 
common units in the ps and the nps. We assume that there are covariates  and responses for everyone
in the two samples, and there are no covariates available for the nonsampled units.  We use the nps (ps) 
to construct a prior  for the ps (nps). We also introduce partial discounting to avoid
a dominance of the prior. We use Bayesian predictive inference for the finite population mean. In our illustrative example on body mass index 
and our simulation study, we compare the relative performance of alternative procedures and demonstrate
that our procedure leads to improved estimates over the ps only estimate.

\medskip

\noindent
keywords:
Covariates, Finite population mean, Inverse probability weighting,
Power prior, Selection bias, Surrogate samples

\newpage

\begin{center}
{\bf 1. Introduction}
\end{center}

Undoubtedly, probability sampling is the gold standard among all data collection procedures. It is
based on randomization, and leads to unbiased and consistent estimates when a proper estimation procedure is implemented. Yet, probability sampling schemes pose difficulties because of high nonresponse rates causing them to lose their much-needed
probabilistic structure. On the other hand, nonprobability samples lack this probabilistic structure and hence
can be grossly inaccurate.  For one thing, they are unlikely to be  representative of the population from which they are 
drawn. However, nonprobability samples are easy to collect and therefore very cheap to run. Faced with expensive
surveys, government agencies are left with no choice but to enormously reduce efforts and costs in planning and fielding probability samples.
Citro (2014) mentioned seven challenges for official statistics and stated,
%Citro (2014) stated “official statistical offices need to move from the probability
%sample survey paradigm for the past 75 years to a mixed model data source
%paradigm for the future”.
``In my view, to respond adequately to one or more, let alone all seven, of these
challenges, official statistical offices need to move from the probability sample survey paradigm of the
past 75 years to a mixed data source paradigm for the future.'' 

It is not clear whether nonprobability sampling can replace probability sampling but the situation
is threatening because of time, cost and nonresponse constraints (e.g., Beaumont, 2020; Rao 2020). Therefore, 
it is important to join scientists in trying to find a resolution to this problem.
Beaumont (2020) reviewed some approaches that can reduce, or even eliminate, the use of probability surveys while
preserving valid statistical inference. All his approaches use data from  nonprobability samples, but in most approaches
probability samples are also used. He particularly discussed the design-based approach based on a probability sample, which is nonparametric, and therefore
is not subject to the risk of bias due to a mis-specified model, but it can be inefficient.
Rao (2020) reviewed various probability survey methods that are used to make valid inferences about finite population parameters.
This allowed him to show how these models can be extended to nonprobability samples that can lead to valid inferences by
themselves or when combined with probability samples.

It is possible to make inference about a finite population quantity using a single nonprobability sample only; see Rao (2020) for a discussion. 
%
%One approach uses poststratification via multi-level regression and poststratification (MRP); see Wang et al. (2015). There are at least three problems with MRP: Sparseness of the
%poststrata sample counts, assumption of known poststrata counts, weak assumption of exchangeability of  poststrata effects. Therefore, it
%is unlikely for MRP to eliminate the selection bias inherent in nonprobability samples; see Valliant (2020) who demonstrated that MRP does not work well for nonprobability samples when there are weak covariates. 
One approach is to use propensity scores to produce nps survey weights and then proceed as in a regular probability sample;  see Elliott and Valliant (2017) for an informative review of quasi-randomization and the super-population approach. Chen, Li and Wu (2020) supplemented a nonprobability sample with a probability sample observing only covariates to estimate propensity scores via logistic
regression. A full Bayesian approach 
in this direction is given by Nandram, Cao, Xu and Bhadra (2019). Another approach is to use a nonignorable selection model to remove the selection bias; see Smith (1983) for pioneering work in this direction. Xu and Nandram (2019) and Xu (2020) used this approach to obtain full Bayesian analyzes.
References in these papers provide a historical development of this area.
It is difficult to  make valid inference from a nonprobability sample with considerable selection
bias.  
%From our investigation, this appears to be correct. 
After all, a probability sample is the gold
standard (high quality), but a nonprobability sample is likely to have low quality (large bias, large mean 
squared error but unrealistically small variance).
% In this paper, within the Bayesian framework, we use 
%both the quasi-randomization approach and the super-population approach.
The key problem of a non-probability sample is that it is very likely to lead to seriously biased
estimates of finite population quantities. Therefore, the large well-documented literature on selection
bias is pertinent in the study of non-probability samples; these articles are too numerous to 
mention here. But see Xu, Nandram and Manandhar (2020) and Choi, Nandram and Kim (2021) for recent applications,
and the references therein. 

It has become necessary to combine the two sampling processes.
There are efforts  to combine both probability and nonprobability samples to produce a
single inference that compensates for the limitations of each process.
Elliott and Haviland (2007) evaluated a composite estimator to supplement
a standard probability sample with a nonprobability sample. They showed
that the estimator, based on a linear combination of both sample processes and a
bias function, can produce estimates with a smaller mean squared error (MSE)
relative to a probability-only sample. 
Elliott (2009) proposed a pseudo-design-based estimation procedure that uses a probability
sample to estimate pseudo-inclusion probabilities for elements of a nonprobability
sample. Both samples are then combined to derive estimates that
are shown to have improved accuracy and smaller MSE compared with those
derived from a probability-only sample; see Elliott and Valliant
(2017) for an informative review. Robbins, Ghosh-Dastidar and Ramchand (2021) proposed
an improved composite estimator, focusing on weights constructed using propensity
scores with consideration given to calibration weighting.
%
%A limitation of these studies is the necessity of a large probability sample to
%produce robust calibration weights or pseudo-inclusion probabilities.
 %
Sakshaug, Wisniowski, Ruiz and Blom (2019), henceforth SWRB, and Wisnioski, Sakshaug, Ruiz and Blom (2020), 
henceforth WSRB, gave very clear comparisons
of a probability sample and a nonprobability sample. SWRB stated: ``Given the advantages of both sampling schemes,
it makes sense to devise a strategy to combine them in a way that is beneficial from both a cost and error perspective.''
In their conclusion, SWRB stated:  ``In conclusion, it is interesting to know that probability and nonprobability
samples can be integrated in a way that exploits their advantages to compensate
for their weaknesses and improve estimation of model parameters.'' 

There are many approaches to making inference about a population using a nonprobability sample only or a nonprobability sample
integrated with a probability sample. In the latter case, it is not clear which should be used to 
supplement the other. SWRB and WSRB, using the Bayesian approach, supported the situation where the nonprobability
sample should be used to supplement the probability sample. We concur with these authors. However, it is possible,
albeit with less quality, a nonprobability sample only can be used with some calibration (benchmarking) of covariates to make  inference about a finite population quantity from the nonprobability sample only; evidently this is risky. Meng (2018) argued that a small bias in big data can be catastrophic; see also Rao (2020) for a review and an interpretation of Meng (2018) relevant to
nonprobability samples.  When the sample is unbalanced with respect to the target population composition,
larger data volume increases the relative contribution of selection bias to absolute or squared error. Meng
(2018) called this phenomenon a ``Big Data Paradox'', and he showed both theoretically and
empirically that the impact of selection bias on the effective sample size can be extremely large.
Primarily, he introduced data defect index, drop out odds and  degree of uncertainty.
As pointed out by  SWRB and WSRB, it will be better to use a nonprobability sample to supplement a probability
sample. 

While both SWRB and WSRB use the Bayesian framework to combine a nonprobability sample and a probability sample,
we believe that this is the right way to go; a description of their method is reviewed in Section 2.1. Indeed, as they correctly pointed out, data integration is a big strength of the
Bayesian framework. Unlike other methods, we can account for variability with virtually no additional effort. 
%In a non-Bayesian
%framework, one would need to appeal to other constructs (e.g.,  variance functions, bootstrapping or asymptotics that understate variability), thereby adding a degree of 
%subjectivity or inconsistency. 
WSRB provided an improved
Bayesian analysis over SWRB. Both SWRB and WSRB are mainly
interested in inference about 
regression coefficients. On the other hand, we are primarily interested in inference about a finite population quantity, especially in the presence
of a nonprobability sample. We combine the nonprobability sample and the probability sample, where one of them is used to
construct the prior for the parameters that are common to both. The question we ask is which one of these two approaches is better. We also show how to 
discount the information provided by the prior in either case. While SWRB did not use survey weights, WSRB
did include survey weights as a covariate. A major part of our work is to incorporate survey
weights into the likelihood (i.e., sampling process).
%We incorporate the survey weights as a product
%of power composite likelihood properly normalized; survey responses are not really independent (especially in the presence
%of survey weights), so when we assume independence, we are actually using a composite likelihood.

Like SWRB and WSRB, we use a multiple linear regression model for the finite population of $N$ units,
$$
y_i \mid \uwd{\beta}, \sigma^2 \stackrel{ind} \sim \mbox{Normal}(\uwd{x}_i^\prime \uwd{\beta}, \sigma^2), 
i=1,\ldots,N,
$$
with appropriate priors on model parameters, $\uwd{\beta}$ and $\sigma^2$, to obtain a full Bayesian approach. 
We assume priors on both parameters that are completely improper and noninformative. That is,
$$
\pi(\uwd{\beta}, \sigma^2) \propto \frac{1}{\sigma^2}, \uwd{\beta} \in R^p, \sigma^2>0.
$$
Models of this form provide proper posterior distributions provided that the matrix of covariates is full rank.
This model holds for all units in the population and so it is called a population model. Within the
Bayesian approach, once a full model is written down, it is not subject to change; nothing else should be
considered when it is fit using MCMC or other sampling-based methods, otherwise it becomes incoherent.
However, for both sampling processes, probability sample and nonprobability sample, we need to
make the necessary adjustments to the population model to accommodate the two processes with survey weights
and discounting; see Pfeffermann (1993) for an adjustment to a population model to account for departures (e.g., selection bias) from the sampling model. Our Bayesian approach is valid for general population models, and 
not just the multiple linear regression model given above. Of course, in any given application we 
will need new specifications; see Appendix A for an example on binary data.

We observe a probability sample  that has survey weights due to the selection of the sample; 
survey weights carry the selection bias that we want to remove. When a non-probability sample is taken,
there are no survey weights and therefore the selection mechanism is not available.
Our main problem is to construct a prior based on the ps (nps) when the actual data come from the nps (ps).
In the nps, the sample size is $n_1$ and in the ps the sample size  is much less,
say $n_2$. In our application, the ps is small and $n_2$ is about 10-20\% of $n_1$.

We note  that it is possible that the nps and ps can have more common covariates 
in the participation model than in the model with the study variable (i.e., there are two sets
of common covariates). We assume that for the participation model, there is a common set of
covariates in the nps and the ps, which we denote by $\uwd{z}$, and  in the study variable model,
there is a second set of common covariates in the nps and ps, which we denote by $\uwd{x}$. 
The set of covariates, $\uwd{x}$, is generally a subset of the set of covariates, $\uwd{z}$.  The population model has the covariates, $\uwd{x}$, it is assumed to be correct, and it is adjusted to accommodate the data of the nps and ps.  The common set of covariates, $\uwd{z}$, are used to estimate the survey weights in the nps. However, in our  illustrative example, these two sets of common covariates are the same. 

%\newpage

There are three problems; see Rao (2020):
\begin{itemize}
   \item[1.] Responses, $y$, and common covariates, $\uwd{x}$, are available in both the ps and the nps. In this case, 
	           one might use either the ps or the
	           nps as the actual sample. 
	 \item[2.] Responses are not observed in the ps. In this case, we might prefer to use the ps to construct 
	           weights for the nps and use the nps as the actual sample. 
	 \item[3.] The covariates, $\uwd{z}$, are observed in the nps and there is no ps. We assume that population totals of 
	           covariates, $\uwd{z}$, are available from a census or administrative records.
	           This can be handled as in Nandram, Cao, Xu and Bhadra (2019),  Xu and Nandram (2019, 2020) and Xu (2020).
\end{itemize}
We study the first problem in this paper. The larger set of common covariates, $\uwd{z}$, may exist in the nps and ps, and these can be used to fit the participation model to help mitigate selection bias.

To focus our development, we study body mass index (BMI) as the variable of interest with covariates, age, race and sex, from eight counties in California, based on a probability sample. Interactions are not included in our models
because they are not statistically significant. The covariates, responses (BMI) and survey weights are all known. 
We construct an example out of these data to obtain a single sample. 
Our construction uses six (6) counties as the nps and the
remaining two (2) as the ps. The weights associated with the nps are discarded, and assumed  to be
unknown. The population is assumed to be set where the ps is taken; 
%so that the population size
%is roughly the sum of the survey weights for the ps. 
the population size is assumed to be roughly the sum of the survey weights for the ps only, and we also 
assume that both the ps and nps are sampled from this population. (The actual sampling fraction is roughly $0.02\%$
in NHANES III.) 
For the study variable, the  covariates and responses in the nps are
 $(\uwd{x}_{1i}, y_{1i}), i=1,\ldots,n_1$.
The survey weights, covariates and responses for the ps are $(W_{2i},\uwd{x}_{2i}, y_{2i}), i=1,\ldots,n_2$.
For the participation model, we have $(W_{1i},\uwd{z}_{1i}), i=1,\ldots,n_1$ for the nps
and $(W_{2i},\uwd{z}_{2i}), i=1,\ldots,n_2$ for the ps; the weights are missing from the nps, but we denote them by 
$W_{1i}, i=1,\ldots,n_1$. Note that in our illustrative example, $\uwd{x}$ and $\uwd{z}$ are exactly the same.
As we stated, this is just an illustrative example. We adjusted a real data set to fit our context, but
real data, which we do not have, are confidential. The data set in SWRB and others have the same form like ours 
though; there are similar confidential data sets in the literature.

Chen, Li and Wu (2020), henceforth CLW, used a ps and an nps to obtain weights for the nps, but no study variable is observed in the ps; only covariates common to ps and nps are observed. The method of CLW is reviewed in Section 2.2. A model for the participation probabilities (propensity scores) for the nps is postulated as a function of the common covariates, $\uwd{z}$, and then estimated by making use of the ps and associated known survey weights. We make use of the inverse of estimated propensity scores as our weights   for the nps. For ready reference, a summary of the Chen, Li and Wu (2020) method of estimating propensity scores for the nps reviewed in Section 2.2. Using the estimated propensity scores, Chen, Li and Wu (2020) proposed inverse probability weighted estimators of the population mean, and established its consistency under the assumed propensity score model for the nps and the design for ps. They also provided variance estimators taking account of the variability in the estimated propensity scores. 

In our application to body mass index (BMI) in Section 4, we have winsorized the estimating weights in both directions; if they are smaller than 1 (not possible under logistic regression model for propensity scores) or if they are extremely large. We also calibrated the winsorized weights to known population totals of common covariates ascertained from web scrapping (i.e., US Census Bureau data on the
internet). Appendix B describes the winsorizing (trimming) and calibration procedure used on the nps weight; see Haziza and Beaumont (2017) for a review of calibration methods.

In our illustrative example, we consider five scenarios (models) to make inference about the finite population mean 
when we have a probability sample and/or a nonprobability sample. It is possible to ignore the nps altogether, and use only the ps. However, if we  have only
the nps, we have to make do with what we have; fielding a small parallel ps may be costly for some agencies. It is possible to
use only the nps, but this has the risk of large bias and the misleading feeling of small variance just because of
its large size. Using only the ps can provide unbiased estimates, but the mean squared error will be large if the sample
size of the ps is small. If the ps is large, there is no use for the nps.
However, the ps gives a good sense of what the point estimate should be. Therefore, when combining the ps and the nps,
although with a relatively small sample size, the ps can at least help to guide the estimation procedure of the nps.

In this method of predicting the finite population mean, there are two models. The first model has the response variable,
called the study variable in nonprobability sampling, is the population model. The other model is the participation model,  which studies the selection indicators. It is this model that provides estimates of the propensity scores
(or selection probabilities) and the reciprocals, calibrated to the population size, are the survey weights.
These survey weights, which are then adjusted, are incorporated into the sampling process via the population model
to get the sample model, where all parameters of the population model are obtained.

As a summary, the novelty of our approach is three-fold. First, we provide a fully Bayesian method to combine two 
likelihoods, one based on the probability sample and the other based on the non-probability sample, incorporating the
survey weights. One of these is used as the prior. Second, the ``prior'' data are partially discounted using a ``power prior'',
thereby preventing the prior data, ps (nps), to dominate the actual data, nps (ps). The power prior is reviewed
in Section 2.3.
Third, we adjust the survey weights to get an effective sample size. This sample size is smaller than the original sample
size, thereby accounting for reduced variability  induced by the design features in drawing the ps. 
These traits make our approach novel, more coherent than SWRB and more 
competitive than WSRB. Except for the discounting factor, which appears only in the data, the parameters are all the same in both the nps and the ps parts of the model. The historical data are used to construct the prior for the
common parameters.

This paper has five sections, including this one, and it is an update of Nandram and Rao (2021). In Section 2, we review preparatory materials in the context of data integration on  a more informative review of Sakshaug et al. (2019),
the method of Chen, Li and Wu (2020) to estimate the propensity scores, and 
the use of the power prior.  In Section 3, we describe our Bayesian methodology that uses discounting and adjusted survey weights. In Section 4, we discuss an application, with some adjustments of the data, on body mass index and we specifically describe the five scenarios (models). We also describe a simulation study to make further comparisons of the five scenarios. Section 5 provides some concluding remarks. The appendices provide technical details on materials such as propensity scores and calibration with some computational details. It also has a bootstrap method to take
care of the variability of the propensity scores when they are incorporated in the sample model, a difficult
problem in the Bayesian paradigm. 

\begin{center}
{\bf 2. Preparatory Materials}
\end{center}

In this section, we provide a brief review of of data integration as presented by Sakshaug et  al. (2019),
method of CLW to estimate the propensity scores (Chen, Li and Wu,2020), and the power prior (e.g., Ibrahim and
Chen 2000).

\begin{center}
{\bf 2.1 A Quick Review of Sakshaug et al. (2019)}
\end{center}

SWRB used a large non-probability sample to supplement a relatively much smaller probability sample (a simple
random sample).
Essentially, under the Bayesian paradigm, they have used the non-probability sample  to provide a prior for the probability sample.
They are interested in super-population parameters, but not really finite populations. 
%Also, survey weights are not studied in their work. 
Therefore, they have studied how well the regression coefficients 
in  the model are estimated. They have also looked at prediction in the sense that they used part 
of the ps to fit the models and predict the part that  is left out. Our work is motivated by SWRB, but we are interested in  prediction for a finite population, a more complex problem than their prediction 
problem. Moreover, we want to integrate both data sets into our likelihood function with appropriate
penalties and adjustments for survey weights.

SWRB assumed that the nps and the ps are fielded by the same questionnaire.
Let $\uwd{y}_2$ denote the vector of $n_2$ responses and $X_2$, a $n_2 \times p$ matrix of covariates associated with the ps, including an
intercept. For their baseline model, called the reference model or Model 1, they assume
$$
\uwd{y}_2 \sim \mbox{Normal}(X_2 \uwd{\beta}, \sigma^2 I).
$$
Apriori, they assume independent priors for the $p$ components of $\uwd{\beta}$,
$$
\beta_j \sim \mbox{Normal}(\beta_{j0}, \sigma^2_{\beta_{j0}}), j=1,\ldots,p
$$
with $\beta_{j0} = 0$ and $\sigma^2_{\beta_{j0}} = 10^6$.
For $\sigma^2$, they assumed $\sigma^2 \sim \mbox{InvGam}(.001, .001)$, a proper diffuse prior. 
Denote the  posterior mean of ${\uwd{\beta}}$ as $\hat{\uwd{\beta}}_2$. Under this non-informative prior, 
the posterior mean makes no use of the nps data.

In Models 2 and 3, they used the non-probability sample to construct informative priors for the probability sample.
Let $\uwd{y}_1$ denote the vector of $n_1$ responses and $X_1$, a $n_1 \times p$ matrix of covariates associated with
the nps, including an
intercept. For Model 2,  they have double-used the ps data through $\hat{\uwd{\beta}}_2$, an incoherent procedure in Bayesian statistics.
They fit a model, similar to Model 1, to the nps to get $\hat{\uwd{\beta}}_1$. Then, they assume informative priors,
$$
\beta_j \sim \mbox{Normal}(\hat{\uwd{\beta}}_{1j}, (\hat{\uwd{\beta}}_{1j}-\hat{\uwd{\beta}}_{2j})^2), j=1,\ldots,p,
$$
but retain the Model 1 prior for $\sigma^2$. 
%Put aside that this prior double-uses the data, the rationale for the chosen
%prior variance of $\beta_j$ is left unclear.
SWRB noted the double use limitation of Model 1 prior and proposed an alternative Model 3 prior to avoid the
double use problem.

In Model 3, SWRB proposed to bootstrap the nps data to obtain a prior for the parameter, $\uwd{\beta}$. They proposed
$$
\beta_j \stackrel{ind} \sim \mbox{Normal}(\hat{\uwd{\beta}}_{2j}, \hat{\sigma}^2_{B,j}), j=1,\ldots,p,
$$
where $\hat{\sigma}^2_{B,j}$ is a bootstrap estimate of variance of $\hat{\uwd{\beta}}_{2j}$. The prior for
$\sigma^2$ is the same diffuse prior used in Models 1 and 2.

%One can believe that Model 2 is not sensible. 
Model 3 indicates a very strong prior on $\beta_j$ because
the bootstrap prior variance above will be very small due to the large size of the nps, and 
is approximately equal to the variance estimate of $\hat{\uwd{\beta}}_{2j}$ obtained by assuming
the regression model holds for the nps. SWRB noted some other limitations of Model 3 prior.
Evidently, this is
unrealistic and suggests one could make inference from the prior only.
Model 1 is fine, but one can simply use the prior $\pi(\sigma^2) \propto 1/\sigma^2$, an objective
prior, and there is no conflict. It is just as simple to fit a model to the nps similar to the
ps. Moreover, clearly independent priors on the regression coefficients are not sensible.

If one needs to use the nonprobability sample to construct a prior for the probability sample, one has to be
careful because this prior can dominate the probability sample. In fact, it actually happens  
in the final estimates. One needs to penalize the prior constructed using the nonprobability
sample.
%
%SWRB could have converted the nps to a ps; weights are not used in SWRB. This can be done by first 
%estimating the survey weights using propensity scores (e.g., Chen, Li and Wu 2020), and then adjusting 
%the  nps to a simple random sample with a smaller effective sample size (i.e. , $n_1$ is now smaller than 
%its original value). It is possible to use surrogate sampling (Nandram 2007) to provide an approximate 
%simple random sample from the population; details are omitted. (An adjustment is not needed for the simple 
%random sample.) Assuming there is no overlap of the two samples, one can fit a model with the same parameters 
%($\uwd{\beta}, \sigma^2$) to both datasets (now both are simple random samples). This model is
%$$
%\uwd{y}_t \mid \uwd{\beta}, \sigma^2 \stackrel {ind} \sim \mbox{Normal}(X_t \uwd{\beta}, \sigma^2 I_{n_t}), t=1,2,
%~~\pi(\uwd{\beta}, \sigma^2) \propto \frac{1}{\sigma^2}.
%$$ 
%%where the length of $\uwd{y}_1$ is $n_1^\ast \leq n_1$, with a corresponding adjustment on $X_1$.
%The joint posterior density of $\uwd{\beta}, \sigma^2$ is proper once at least one of
%$X_1$ or $X_2$ is full rank.

Based on the simulation study they did for both estimation and prediction, the results appear very good;
prediction results being less convincing. The results indicate that an increase in bias due to using the
nps-based prior, is offset by a reduction in variance, but a degree of bias still remains. 
%They found that Model 3 is the best in this respect. 
SWRB also performed an analysis 
on a real dataset and the results appear satisfactory. 
%In the discussion, they stated:
%``In conclusion, we find that augmenting a probability sample with a nonprobability sample under the
%Bayesian framework can produce survey estimates with smaller mean squared error and potentially 
%large cost savings relative to probability-only samples.''   

%\newpage
\begin{center}
{\bf 2.2 A Review of the CLW Method for Estimating Propensity Scores}
\end{center}

Let $\uwd{z_{i}},i=1,\ldots,N$, denote the common (nps and ps) covariates used in the  participation model. These are observed in the ps and the nps,
but not observed for the rest of the population. Again, for the nps, we have $\uwd{z}_{1i}, i=1,\ldots,n_1$,  and
for the ps, we have $\uwd{z}_{2i}, i=1,\ldots,n_2$. Chen, Li and Wu (2020) have a method with two key ideas
to estimate the propensity scores for the nps, and therefore the survey weights that are proportional to the 
reciprocals of the propensity scores. They assume that the propensity scores can be modeled parametrically
using
$$
\pi_i = P(R_i=1 \mid \uwd{z}_i) = \pi(\uwd{z}_i; \uwd{\theta}),
$$
with independence over $i$, where $\uwd{\theta}$ are to be estimated. Here $R_i=1$ for the ps or nps; $R_i = 0$ for the nonsamples. Then, the population likelihood function is
$$
\ell(\uwd{\theta}) = \prod_{i=1}^N \{\pi(\uwd{z}_i; \uwd{\theta})\}^{R_i}\{1-\pi(\uwd{z}_i; \uwd{\theta})\}^{1-R_i}.
$$

The first key idea is to write the log-likelihood as
$$
\ell_1(\uwd{\theta}) = \sum_{i=1}^{n_1} \log \left \{ \frac{\pi(\uwd{z}_{1i}; \uwd{\theta})}{1-\pi(\uwd{z}_{1i}; \uwd{\theta})} \right \}
+ \sum_{i=1}^N \log \{1-\pi(\uwd{z}_i; \uwd{\theta})  \}.
$$
The second key idea is to replace the unknown last term in $\ell_1(\uwd{\theta})$ by its unbiased estimator (Horvitz-Thompson) based
on the ps and associated design weights $W_{2i}$ since the nonsample $\uwd{z}_i$ are unknown. The resulting pseudo-log-likelihood is given by
%use the pseudo-log-likelihood by replacing the second term by the Horvitz-Thompson 
%estimator since the nonsample $\uwd{x}_i$ are unknown, as
$$
\ell_1(\uwd{\theta}) = \sum_{i=1}^{n_1} \log \left \{ \frac{\pi(\uwd{z}_{1i}; \uwd{\theta})}{1-\pi(\uwd{z}_{1i}; \uwd{\theta})} \right\}
+ \sum_{i=1}^{n_2} W_{2i} \log \{1-\pi(\uwd{z}_{2i}; \uwd{\theta}) \},
$$
which can now be maximized with respect to $\uwd{\theta}$ to get a pseudo maximum likelihood estimator, 
$\hat{\uwd{\theta}}$. The propensity scores for the nps are then
$\pi(\uwd{z}_{1i}; \hat{\uwd{\theta}}), i=1,\ldots,n_1$. Henceforth, Chen, Li and Wu (2020) specialize to logistic regression.

The gradient vector of the pseudo-log-likelihood in the general form is
$$
\Delta(\uwd{\theta}) = \sum_{i=1}^{n_1} \{1-\pi(\uwd{z}_{1i}; \uwd{\theta})\}
\frac{\partial \pi(\uwd{z}_{1i}; \uwd{\theta})}{\pi(\uwd{z}_{1i}; \uwd{\theta})}
-\sum_{i=1}^{n_2} W_{2i} \frac{\partial \pi(\uwd{z}_{2i}; \uwd{\theta})}{1-\pi(\uwd{z}_{2i}; \uwd{\theta})},
$$
where $\partial \pi(\uwd{z}_{1i}; \uwd{\theta})$ or $\partial \pi(\uwd{z}_{2i}; \uwd{\theta})$ is the gradient vector 
with respect to $\uwd{\theta}$.
Then, $\Delta(\hat{\uwd{\theta}}) = \uwd{0}$ give the solutions. In the case of logistic regression, Chen, Li and Wu (2020) have used the Newton-Ralphson method
to solve the equations, starting with $\hat{\uwd{\theta}} = \uwd{0}$, but we note that the Newton-Raphson's method
is sensitive to those starting values. It is also computationally unstable for small samples.
%
%It is worth noting that Chen, Li and Wu (2020) did not incorporate the uncertainty in the estimation
%of propensity scores into their methodology when the propensity scores are used for prediction of the 
%finite population mean; see Nandram, Cao, Xu and Bhadra (2019) for  a Bayesian method that adjusts for this
%uncertainty. However, in the case of Chen, Li and Wu (2020) further study is needed.

\begin{center}
{\bf 2.3 Power Prior}
\end{center}

The power prior is an informative prior that combines historical data with current data. It has been extensively used in
many different applications over the past thirty years. However, we must be careful that the historical data do not 
dominate the current data. In an interesting paper, Ibrahim and Chen (2000) reviewed the properties of the power prior
for arbitrary regression models (e.g., linear models and generalized linear models). They discussed many applications
in diverse disciplines. However, they expressed doubts about the computational aspect of the power prior. Later
Ibrahim, Chen, Gwon and Chen (2015) reviewed many theoretical properties for a much larger class of models with many different
formulations. There are many versions of the power prior. 

In our application, we can treat the probability sample
or the non-probability sample as historical data with the appropriate power prior. The power prior is a penalty to the
historical data partially discounting it; here the ps or the nps is the historical data. 
%One of the shortcomings of the SWRB work is that it does not penalize the non-probability sample when they used it to %supplement the probability sample. 
We show how to use the power prior to partially discount the non-probability sample and vice versa.

Suppose $y \mid \uwd{\theta} \sim f(y \mid \uwd{\theta})$. Then, one version of the the power prior is
$$
f(y \mid \uwd{\theta}, a) = \frac{ \{f(y \mid \uwd{\theta}) \}^a}
{\int_{y} \{f(y \mid \uwd{\theta})\}^a dy}, 0 \leq a \leq 1.
$$
It is important to note that $a$  may not be identifiable in this power prior alone.

If the historical data, $y_{11},\ldots,y_{1 n_1}$, and current data,   $y_{21},\ldots,y_{2 n_2}$, are available, then assuming that the historical data and current data are independent, the
joint probability `density' function  of the data is
$$
f(\uwd{y}_1, \uwd{y}_2 \mid \uwd{\theta}, a)
= \prod_{i=1}^{n_1} \frac{ \{f(y_{1i} \mid \uwd{\theta}) \}^a}
{\int_{y_{1i}} \{f(y_{1i} \mid \uwd{\theta})\}^a dy_{1i}}
\prod_{i=1}^{n_2} f(y_{2i} \mid \uwd{\theta}), 0 \leq a \leq 1.
$$ 
Here $a$ may be identifiable, but some control may be needed over $a$ in general. Besides for generalized
linear models, there may be difficulties in computation, particularly the normalization constant. This problem is particularly  severe when Markov chain Monte Carlo methods are used for computations. This prompted Ibrahim, Chen, Gwon and Chen (2015) and others to specify $a$, and they suggested many methods for doing so.  This leads to extensive sensitivity analysis. We keep $a$ random, and if it is needed, $a$ can be specified to be  in a sub-interval of $(0, 1)$ (e.g., $(.5, 1)$ for at most 50\% discounting) rather than to actually specify $a$. Besides it is  really a bad idea to specify $a$ in a Bayesian setting because $a$ serves a very important function in the model; the data should `speak' for it.

Let us consider a simple example for the power prior,
%$$
%y_{11},\ldots,y_{1 n_1} \mid \theta, \sigma^2 \stackrel{iid} \sim \mbox{Normal}(\theta, \sigma^2).
%$$
%Then, for the power prior,
$$
y_{11},\ldots,y_{1 n_1} \mid \theta, \sigma^2, a \stackrel{iid} \sim \mbox{Normal}(\theta, \frac{\sigma^2}{a})
$$
and for the current data,
$$
y_{21},\ldots,y_{2 n_2} \mid \theta, \sigma^2 \stackrel{iid} \sim \mbox{Normal}(\theta, \sigma^2).
$$
Clearly, $a$ is not identifiable in the power prior if $\sigma^2$ is also unknown. However, if $\sigma^2$
is known, $a$ serves the role as a penalty to increase variance (i.e., $\sigma^2$ to $\sigma^2/a, 0<a<1$).

Assume the prior $\pi(\theta, \sigma^2, a) \propto 1/\sigma^2$.
Letting $D = (\uwd{y}_1, \uwd{y}_2)$ and using Bayes' theorem,
the joint posterior density is
$$
\pi(\theta, \sigma^2, a \mid D) \propto
a^{n_1/2} \left(\frac{1}{\sigma^2}\right)^{(n_1+n_2)/2+1}  \times
$$
$$
\exp \left \{-\frac{1}{2\sigma^2}\{a(n_1-1)s_1^2 + (n_2-1)s_2^2   
+ an_1(\bar{y}_1-\theta)^2 + n_2(\bar{y}_2-\theta)^2\} \right\}, 0 \leq a \leq 1,
$$
where $\bar{y}_t$ $s_t^2, t=1, 2$, are the sample means and the sample variances.

Letting  $\lambda = \frac{an_1}{a n_1 + n_2}$, it follows that
$$
\theta \mid \sigma^2, a, D \sim \mbox{Normal}\{\lambda \bar{y}_1 + (1-\lambda) \bar{y}_2, (1-\lambda) \sigma^2/n_2\},
$$
$$
\sigma^2 \mid a, D \sim \mbox{InvGam}\left\{\frac{n_1+n_2-1}{2},  \frac{n_2\lambda (\bar{y}_1-\bar{y}_2)^2 + a(n_1-1)s_1^2 + 
(n_{2}-1)s_2^2}{2} \right\}
$$
and
$$
\pi(a \mid D) \propto \frac{a^{n_1/2} \sqrt{(1-\lambda)/n_2}} { \{(n_2\lambda (\bar{y}_1-\bar{y}_2)^2 + a(n_1-1)s_1^2 + 
(n_{2}-1)s_2^2\}^{(n_1+n_2-1)/2}}, 0 \leq a \leq 1.
$$
Therefore, $\pi(a \mid D)$ is well defined for all $0 \leq a \leq 1$ and $a$ is properly identifiable. Moreover, there will be no difficulties in
computation because $a$ can be sampled using a grid method.

The difference between this illustration and our problem is that we have survey weights and covariates. But the
implementation is similar in that we do not need to use a Gibbs sampler; we can use the multiplication rule to 
draw the samples as in this illustrative example. As pointed out in the literature on power priors, it is possible to have
poor mixing when Markov chain Monte Carlo methods (e.g., the Gibbs sampler) are used; this is caused by the 
stochastic feature in $a$ and this is why researchers have turned away from a full specification of $a$ together with 
extensive sensitivity analysis, a nuisance.  But this slow mixing can be avoided by using a carefully planned 
block Gibbs sampler instead (not need and not studied here).

It is worth noting that the discounting factor is used to penalize the prior data, nps or ps. It has nothing 
to do with selection bias; it is only the survey weights that account for selection bias, nothing else. The discounting 
factor is used in the historical data (nps or ps) to construct a prior for the parameters of the model in such a way
that the prior does not dominate the likelihood (i.e., if the likelihood is from the ps, the nps is used as historical data and vice versa). Here, the parameters $\theta$ and $\sigma^2$ are the same in both data sets, and the historical
data are used to construct a prior for $(\theta, \sigma^2)$.
This is particularly important when the nps is used as historical data. The prior for the discounting factor, $a$, is a uniform distribution on $(0,1)$ that is proper but essentially noninformative (i.e., the likelihood remains invariant
under its use).

%%%%%%%%%%%%%%%%%%%%%%%%%%%%%%%%%%%%%%%%%%%%%%%%%%%%%%%%%%%%%%%%%%%%%%%%%%%%%%%%%%%%%%%%%%%%%%%%%%%%%%%%%%%%%%%%%%%
%%%%%%%%%%%%%%%%%%%%%%%%%%%%%%

\begin{center}
{\bf 3. Bayesian Methodology}
\end{center}

We consider the homogeneous case (i.e., no subgroups or clustering), consisting of a nonprobability sample and a probability sample.
In the construction here, the survey weights are very important and they are to be used to avoid bias due to the
sample design and other post-design adjustments such as adjustment for nonresponse. 
However, it is worth mentioning again that we are considering the case of a homogeneous sample (population). That is,
there are no sub-groups (e,g., small areas) or clustering. We also discuss how to partially discount the prior
data (ps or nps).

We note here that the nps and the ps are assumed to be independent samples from the same population. The ps is a probability sample, and therefore, it is a representative sample. The nps is
a nonprobability sample, and therefore, it may not be a representative sample from the same population. We estimate propensity
scores (survey weights) to make the nps compatible with the ps; see Appendix A. If nps (ps) is used
as a prior, there is discounting of the nps (ps) because the prior data may be regarded as historical data.

First, in our approach we need the effective sample size and the adjusted survey weights.
Let $y_i, i=1,\ldots,n$, be independent with E$(y_i) = \mu_i$,  var$(y_i) = \nu_i^2$ and $W_i,i=1,\ldots,n$ be the
original survey weights. This is a 
standard assumption in a
super-population model, but it is questionable as the units may not be independent.
Then, with just this assumption,
Potthoff, Woodbury  and Manton (1992) showed that the equivalent (effective) sample size is ${n}_o$, where
$$
{n}_o = \frac{(\sum_{i=1}^n W_i)^2} {\sum_{i=1}^n W_i^2}.
$$
The effective sample size, ${n}_o$, indicates the extent to which the variance is increased by the unequal weighting;
see Kish (1965) for design effect.
Then, the adjusted survey weights required to eliminate bias introduced by the original survey weights are
\begin{equation}
\label{aw}
w_i = {n}_o \frac {W_i}{\sum_{j=1}^n W_j}, i=1,\ldots,n.
\end{equation}
[Note the use of small $w$ for adjusted survey weights.]
Here $\sum_{i=1}^n W_i = N$, the population size, and $\sum_{i=1}^n w_i = {n}_o = \sum_{i=1}^n w_i^2$.
We note that ${n}_o$ has some interesting properties. First, if the $W_i$ are nearly equal, ${n}_o = n$. 
Second, if $n >1$, ${n}_o > 1$.
Third, ${n}_o$ is invariant to scale and therefore the $w_i$ are invariant to scale.
The adjusted weights in (\ref{aw}) will play an important role in the Bayesian methodology. 
%In Appendix D, we
%provide an effective sample size when the independence assumption might not hold.

In Section 3.1, we discuss the Bayesian models, which are used to integrate the nps and the ps.
In Section 3.2, we discuss Bayesian predictive inference using surrogate sampling.
In Section 3.3, we make some additional observations about selection bias, discounting and
robustness.

\begin{center}
{\bf 3.1 Bayesian Models}
\end{center}

To introduce the Bayesian models, we start with the simple case of the finite population with values
obeying the model,
$y_1,\ldots,y_N \mid \uwd{\theta} \stackrel{iid} \sim f(y \mid \uwd{\theta})$.
Therefore, we can now write the joint density of $y_1,\ldots,y_n$ as a weighted product,
\begin{equation}
\label{bd}
g(\uwd{y}\mid \uwd{\theta}, \uwd{w}) = \prod_{i=1}^n \frac{\{f(y_i \mid \uwd{\theta})\}^{w_i}}
{\int \{f(y_i \mid \uwd{\theta})\}^{w_i}dy_i},
\end{equation}
where we have conditioned on $\uwd{w}$ to cover the case when the survey weights are random. [Henceforth, we
will drop the conditioning on $\uwd{w}$.] Apart from the normalization constant, this is similar to what we do in survey sampling. However, generally the denominator in (\ref{bd}) is important. 
%and the normalization constant will make no difference under normality.

In (\ref{bd}), we are assuming that the population is homogeneous. If the population is heterogeneous, and we know the
sub-groups (e.g., small areas or clusters), (\ref{bd}) must be applied to each of them separately. Also, the effective
sample size formula given here is applicable under independence. 

Second, we describe how to use the power prior to partially discount for the prior data.
If we  put a prior on $\uwd{\theta}$, say $\pi(\uwd{\theta})$, using Bayes' theorem, the posterior distribution of
$\uwd{\theta}$ is
$$
\pi(\uwd{\theta} \mid \uwd{y}) \propto \pi(\uwd{\theta})g(\uwd{y} \mid \theta),
$$
which we assume is proper. Actually, we want to construct an informative prior for $\uwd{\theta}$ using the
nps (ps) when the actual data are the ps (nps). We want to do so to avoid the prior from dominating the
actual sample information, and therefore, some discounting is necessary, especially when the nps is used to construct the prior. That is, the nps (ps) is used to construct a data-based prior with
some discounting; the ps and nps are assumed independent. The power prior can be used for this purpose;
see Ibrahim and Chen (2000) and Ibrahim, Chen, Gwon and Chen (2015) again.

The nps or ps can be used separately, but our main contribution is to show how to combine 
these two samples. If the nps is used as a prior, we need to penalize it because it may be 
biased with small variance and it can dominate the ps. If we have only the nps data, there 
is no way to penalize it. Therefore, if only the nps is used, we will set $a=1$, and if it
is used as a prior for ps, $a$ will be random in $(0, 1)$.

It is important to note that the adjusted survey weights, $w_{1i},i=1,\ldots,n_1$, for the nps 
are estimated  using the CLW method. Here the weights and covariates from the ps are coupled with
the covariates from the nps to get the propensity scores using logistic regression. It is also 
worth noting the study variables in both the ps
and nps are not used; unlike our work, CLW did not assume the existence of the study variable 
$y$ for the ps. These estimated weights are assumed known in the Bayesian analysis, where we model 
the study variables, not the survey weights. We now turn to the case where the population values,
$y_1,\ldots,y_N$, are assumed to be independent with density function, $f(y_i\mid \uwd{x}_i,\uwd{\theta})$,
where $\uwd{x}$ is a possible subset of  $\uwd{z}$ and $\uwd{\theta}$ a vector of parameters.

First, consider the case when the nps is used on its own with $a=1$, and the special case of linear 
regression, $y_1,\ldots,y_N \mid \uwd{\beta}, \sigma^2 \sim \mbox{Normal}(\uwd{x}_i^\prime\uwd{\beta}, \sigma^2)$
with parameters, $\uwd{\theta} = (\uwd{\beta},\sigma^2)$,
where we have used the prior $\pi(\uwd{\beta}, \sigma^2) \propto 1/\sigma^2$.
Then, the joint posterior density of $(\uwd{\beta},\sigma^2)$, after adjusting for possible
selection bias through the propensity score weights $w_{1i}$, is given by
$$
\pi(\uwd{\beta}, \sigma^2 \mid \uwd{y}_1, a=1) \propto \frac{1}{\sigma^2}
\prod_{i=1}^{n_1} \{\frac{aw_{1i}}{\sigma^2}\}^{1/2}
e^{-\frac{a}{2\sigma^2} \sum_{i=1}^{n_1} w_{1i}(y_{1i}-\uwd{x}_{1i}^\prime \uwd{\beta})^2}.
$$
It is easy to draw samples of $(\uwd{\beta}, \sigma^2)$ because
$$
\uwd{\beta} \mid \sigma^2, a=1, \uwd{y}_1 \sim \mbox{Normal} \left \{\hat{\uwd{\beta}}, \frac{\sigma^2}{a} \left(
\sum_{i=1}^{n_1} w_{1i} \uwd{x}_{1i} \uwd{x_{1i}}^\prime \right)^{-1} \right\},
$$
where $\uwd{\hat{\beta}} = \left(\sum_{i=1}^{n_1} w_{1i} \uwd{x}_{1i} \uwd{x_{1i}}^\prime \right)^{-1} 
\sum_{i=1}^{n_1} w_{1i} \uwd{x}_{1i} y_{1i}$ and
$$
\sigma^2 \mid a=1,\uwd{y}_1 \sim \mbox{InvGam}\{ \frac{n_1-p}{2},   ~~\frac{a \sum_{i=1}^{n_1} w_{1i}(y_{1i}-\uwd{x}_{1i}^\prime \uwd{\hat{\beta}})^2 }{2}  \}.
$$
This method, based only on the nps, is used for comparison. It may be noted that the population model
may not hold for the nps if the common covariates, $\uwd{x}$, used in the population model, is a subset of the 
common covariates $\uwd{z}$, used in the participation model.

Second, we consider the case when the nps is used to construct the prior for the ps. In this case,
$$
\pi(\uwd{\beta}, \sigma^2, a \mid \uwd{y}_1) \propto \frac{1}{\sigma^2}
\prod_{i=1}^{n_1} \{\frac{aw_{1i}}{\sigma^2}\}^{1/2}
e^{-\frac{a}{2\sigma^2} \sum_{i=1}^{n_1} w_{1i}(y_{1i}-\uwd{x}_{1i}^\prime \uwd{\beta})^2},
$$
where $\pi(\uwd{\beta}, \sigma^2, a) \propto 1/\sigma^2$ with $a \sim \mbox{Uniform}(0,1)$.
Note that $a$ is not identifiable in this prior because $\frac{\sigma^2}{a}$ is identifiable but 
not $a$ or $\sigma^2$ separately. That is, the prior is improper, but this does not matter
because the joint posterior density is proper; see Appendix C.
 %
%However,
%as $a$ is not identified, one can perform a sensitivity analysis, but this is not really
%of any importance. It is really the posterior density after both samples are incorporated
%that is of greatest interest.

For our problem of data integration, if the nps is used to construct the prior and the ps is the 
actual sample, the joint posterior density is
$$
\pi(\uwd{\theta}, a \mid \uwd{y}_1, \uwd{y}_2) \propto
\pi(\uwd{\theta}) \pi(a)
 \prod_{i=1}^{n_1} \frac{\{f(y_{1i} \mid \uwd{\theta})\}^{aw_{1i}}}
{\int \{f(y_{1i} \mid \uwd{\theta})\}^{aw_{1i}}dy_{1i}}
\prod_{i=1}^{n_2} \frac{\{f(y_{2i} \mid \uwd{\theta})\}^{w_{2i}}}
{\int \{f(y_{2i} \mid \uwd{\theta})\}^{w_{2i}}dy_{2i}}, 0 \leq a \leq 1,
$$
where $a$ is a discounting factor for the power prior. Similarly, if 
the ps is used to construct the prior and nps is the actual sample,
the joint posterior density is
$$
\pi(\uwd{\theta}, a \mid \uwd{y}_1, \uwd{y}_2) \propto
\pi(\uwd{\theta}) \pi(a)
 \prod_{i=1}^{n_1} \frac{\{f(y_{1i} \mid \uwd{\theta})\}^{w_{1i}}}
{\int \{f(y_{1i} \mid \uwd{\theta})\}^{w_{1i}}dy_{1i}}
\prod_{i=1}^{n_2} \frac{\{f(y_{2i} \mid \uwd{\theta})\}^{aw_{2i}}}
{\int \{f(y_{2i} \mid \uwd{\theta})\}^{aw_{2i}}dy_{2i}}, 0 \leq a \leq 1.
$$
We note that although $a$ is not identifiable in the prior alone, as the two
samples are combined, $a$ becomes identifiable because $\sigma^2$ can be estimated
using only the ps. We will demonstrate this below.

We assume no overlaps between the nps and ps. That is, a unit is not captured in both
the ps and the nps, a standard assumption. We note  that to stay within the Bayesian paradigm, we cannot 
use the data on the study variable, $y$, to construct the prior; all other variables can be used. 
%We can not use both
%samples for inference because we do not know the identity of the units in the nps;  we run the
%risk of double counting a unit.

%In this case, we will construct the weights from the ps to get the weights in nps.
%We will calibrate (benchmark) the nps covariates to the total from the nps covariates or from a census or
%administrative records (or web scraping).
%Fill in the weights in the nonprobability sample, as we just described. However, we  need the
%covariates in the nps because we do need to make posterior inference for a finite population quantity.
%So we can construct the prior from the nps. However, we do need the nonsample covariates in the
%probability sample, although this can be overcome using surrogate sampling (e.g., Bhatta, Nandram and Sedransk 2018).
 %Therefore, we can use the covariates in both the nps (ps) to obtain the nonsampled 
%covariates in the ps (nps).

For the linear regression case, with the nps as prior, the appropriate posterior density is
$$
\pi(\uwd{\beta}, \sigma^2, a \mid \uwd{y}_1,\uwd{y}_2) \propto
$$
$$
\frac{1}{\sigma^2}
\prod_{i=1}^{n_1} \{\frac{aw_{1i}}{\sigma^2}\}^{1/2}
e^{-\frac{a}{2\sigma^2} \sum_{i=1}^{n_1} w_{1i}(y_{1i}-\uwd{x}_{1i}^\prime \uwd{\beta})^2}
\prod_{i=1}^{n_2} \{\frac{w_{2i}}{\sigma^2}\}^{1/2}
e^{-\frac{1}{2\sigma^2} \sum_{i=1}^{n_2} w_{2i}(y_{2i}-\uwd{x}_{2i}^\prime \uwd{\beta})^2}, 0 \leq a \leq 1,
$$
where we assume a noniformative uniform prior on $a$. Here, the model parameters can be drawn from the posterior 
distribution by simply using
a random sampler (not a Markov chain). It is possible to integrate out $\uwd{\beta}$ and $\sigma^2$
to get the posterior density of $a$; and samples from the posterior density of $a$ can be drawn easily using a grid method. Also, the posterior propriety can be established.

Letting $\uwd{y} = (\uwd{y}_1, \uwd{y}_2)$, the joint posterior density can be written as
$$
\pi(\uwd{\beta}, \sigma^2, a \mid \uwd{y}) \propto a^{n_1/2} (\frac{1}{\sigma^2})^{\frac{n_1+n_2}{2}+1}
e^{-\frac{1}{2\sigma^2} Q}, 0 \leq a \leq 1,
$$
where $Q = a \sum_{i=1}^{n_1} w_{1i}(y_{1i}-\uwd{x}_{1i}\uwd{\beta})^2 + 
\sum_{i=1}^{n_2} w_{i2}(y_{2i}-\uwd{x}_{2i}\uwd{\beta})^2.$ In Appendix C, we obtain a random sampler
to draw from the joint posterior density of $\uwd{\beta}, \sigma^2, a \mid \uwd{y}$ and we show that the joint posterior density is proper.
We state the main distributions to get a random sampler.

For convenience, letting $a_1=a, a_2=1$ (i.e., nps is used as a prior), we define
$$
A = \sum_{s=1}^2 \sum_{i=1}^{n_s} a_s w_{si}\uwd{x}_{si} \uwd{x}_{si}^\prime, 
~\uwd{b} =   \sum_{s=1}^2 \sum_{i=1}^{n_s} a_s w_{si}\uwd{x}_{si} y_{si}
~\mbox{and}~ d = \sum_{s=1}^2 \sum_{i=1}^{n_s}a_s w_{si} (y_{si}-\uwd{x}_{si}^\prime \hat{\uwd{\beta}})^2.
$$
Then, letting $\hat{\uwd{\beta}} = A^{-1} \uwd{b}$, we have
$$
\uwd{\beta} \mid \sigma^2, a, \uwd{y} \sim \mbox{Normal}(\hat{\uwd{\beta}}, \sigma^2 A^{-1}),
$$
$$
\sigma^2 \mid a, \uwd{y} \sim \mbox{InvGam}\left(\frac{n_1+n_2-p}{2}, \frac{d}{2} \right),
$$
and
$$
\pi(a \mid \uwd{y}) \propto \frac{a^{n_1/2} \mid A \mid^{-1/2}}{d^{(n_1+n_2-p)/2}}, 0 \leq a \leq 1.
$$
The joint posterior density is proper because $0 \leq a \leq 1$ and all quantities are well defined,
 provided the design matrix, $X_1$ (or $X_2$), of the nps (ps)  
is full rank. We can draw a sample from $\pi(a \mid \uwd{y})$ 
using the grid method. The other parameters, $\uwd{\beta}$ and $\sigma^2$, are drawn in a standard
manner. However, the main inconvenience is to find the inverse and the determinant of $A$ at each
value of $a$ because $a$ is jittered at each draw from its discrete distribution (i.e., the
draws of $a$ are different almost surely). 
%It is possible to get around this difficulty using a 
%sub-sampling procedure, albeit with less quality; yet we might want to do so.

Here, we assume that the actual sample is the ps, and the nps plays the primary role in the construction
of the prior. By interchanging the roles of $a_1$ and $a_2$, we
will get the posterior density for the case where nps is the actual sample and the ps plays the primary 
role in the construction of the prior.

\begin{center}
{\bf 3.2 Bayesian Predictive Inference}
\end{center}

It is difficult to do prediction fully within the Bayesian paradigm because the nonsampled values of the covariates 
are unknown and the population is fairly large. One can fill in the nonsampled covariates subject to some 
constraint like the total for each covariate is known. In this case, one would need to sample the entire 
population using surrogate sampling. That is, once the parameters are estimated from the sample model 
(population model adjusted with survey weights and the discounting factor), one can then use the appropriate parameters 
in the population model to sample all the $N$ values of the study variable (i.e., projective inference). 
This is surrogate sampling (see Nandram 2007, Nandram  and Choi 2010 and many others). However, for a large 
population this is a time-consuming procedure that depends on external sources of information (e.g., 
administrative records for $N$ and the total for each covariate).
% Based on quasi randomization, we have used an alternative procedure.
%Based on randomization, we have made an adjustment to surrogate sampling.

Under the population linear regression model,
$$
y_i \mid \uwd{\beta},\sigma^2, \uwd{x}_i \stackrel{ind} \sim \mbox{Normal}(\uwd{x}_i^\prime \uwd{\beta}, \sigma^2),
i=1,\ldots,N,
$$
we have the finite population,
\begin{equation}
\label{pm}
\bar{Y} \mid \uwd{\beta}, \sigma^2 \sim \mbox{Normal}(\bar{\uwd{X}}^\prime \uwd{\beta}, \frac{\sigma^2}{N}),
\end{equation}
where $\bar{X} = \frac{1}{N}\sum_{i=1}^N \uwd{x}_i$ and $\bar{X}$ and  $N$ may be unknown.
Note that there are no survey weights in $f(\bar{Y} \mid \uwd{\beta}, \sigma^2)$ in (\ref{pm}).
If we know $\bar{X}$, $N$,  $\uwd{\beta}$ and $\sigma^2$, we can make inference about the finite population mean, $\bar{Y}$.
However, within the Bayesian paradigm, we should really use the posterior density,
\begin{equation}
\label{pred}
f(\bar{Y} \mid \uwd{y}_1, \uwd{y}_2) = \int f(\bar{Y} \mid \uwd{y}_1, \uwd{y}_2, \uwd{\beta}, \sigma^2)
\pi(\uwd{\beta},\sigma^2 \mid \uwd{y}_1, \uwd{y}_2) d\uwd{\beta} d\sigma^2.
\end{equation}
All our efforts in data integration went into building the posterior density, $\pi(\uwd{\beta},\sigma^2 \mid \uwd{y}_1, \uwd{y}_2)$. 

Note that $f(\bar{Y} \mid{y}_1,\uwd{y}_2,\uwd{\beta}, \sigma^2)$ in (\ref{pred}) has
no survey weights and both $\uwd{y}_1$ and $\uwd{y}_2$ are corrupted. 
This is why
we need to use surrogate sampling to replace 
$f(\bar{Y} \mid \uwd{y}_1, \uwd{y}_2, \uwd{\beta}, \sigma^2)$
by  $f(\bar{Y} \mid  \uwd{\beta}, \sigma^2)$ in (\ref{pred}). This is a form of Bayesian projective (not predictive)
inference, and the entire population is sampled without selection bias (i.e., surrogate sampling).

%
%Let $\bar{y}_{2s} = \frac{1}{n_2} \sum_{i=1}^n y_{2i}$, the ps average response, $\bar{\uwd{x}}_{2s} = \frac{1}{n_2} \sum_{i=1}^n \uwd{x}_{2i}$, the ps average covariate, $\bar{\uwd{X}} = \frac{1}{N}\sum_{i=1}^N \uwd{x}_i$, population average covariate, $N$, the population size, and $f_2=\frac{n_2}{N}$, the sampling fraction. Then, it is true that given $\uwd{\beta}, \sigma^2$ and  $\uwd{y}_{2s}$,
%$$
%\bar{Y}  \mid \mathunderaccent\tilde{\beta}, \sigma^2, \mathunderaccent\tilde{y}_{2s}
%\sim \mbox{Normal}\left( f_2(\bar{y}_{2s} - \bar{\uwd{x}}_{2s}^\prime\uwd{\beta}) +\bar{\uwd{X}}^\prime
%\mathunderaccent\tilde{\beta}, (1-f_2)\frac{\sigma^2}{N}\right),
%$$
%$$
%\bar{Y}  \mid \mathunderaccent\tilde{\beta}, \sigma^2, \mathunderaccent\tilde{y}_s
%\sim \mbox{Normal}\left(\bar{\uwd{X}}^\prime
%\mathunderaccent\tilde{\beta}, \frac{\sigma^2}{N}\right),
%$$
%Note that if $n<<N$ (i.e, $f_2 \approx 0$), 
%$\bar{Y}  \mid \mathunderaccent\tilde{\beta}, \sigma^2, \mathunderaccent\tilde{y}_{2s}
%\sim \mbox{Normal}\left( \bar{\uwd{X}}^\prime
%\mathunderaccent\tilde{\beta}, \frac{\sigma^2}{N}\right)$,
%and the $\bar{Y}  \mid \mathunderaccent\tilde{\beta}, \sigma^2, \mathunderaccent\tilde{y}_{2s}$
%does not depend on $\mathunderaccent\tilde{y}_{2s}$.

Because $\bar{\uwd{X}}$ and $N$ 
are both unknown, in the Bayesian paradigm, these are parameters. Here, using the ps  and 
inverse probability weighted estimators, we approximate (\ref{pm}) by
%$$
%\bar{Y} \mid \left \{\mathunderaccent\tilde{\beta}, \sigma^2, 
%\bar{\uwd{X}} = \frac{\sum_{i=1}^{n_2} W_{2i} \mathunderaccent\tilde{x}_{2i}}{\sum_{i=1}^{n_2} W_{2i}}, N = \sum_{i=1}^{n_2} W_{2i}, \mathunderaccent\tilde{y}_s \right \}
%\sim 
%$$
%$$
%\mbox{Normal}\left\{ f_2(\bar{y}_{2s} - \bar{\uwd{x}}_{2s}^\prime\uwd{\beta}) + \frac{\sum_{i=1}^{n_2} W_{2i} \mathunderaccent\tilde{x}_{2i}^\prime}{\sum_{i=1}^{n_2} W_{2i}}
%\mathunderaccent\tilde{\beta}, (1-f_2)\frac{\sigma^2}{\sum_{i=1}^{n_2} W_{2i}}\right\},
%$$
$$
\bar{Y} \mid \left \{\mathunderaccent\tilde{\beta}, \sigma^2, 
\bar{\uwd{X}} = \frac{\sum_{i=1}^{n_2} W_{2i} \mathunderaccent\tilde{x}_{2i}}{\sum_{i=1}^{n_2} W_{2i}}, N = \sum_{i=1}^{n_2} W_{2i}  \right \}
\sim 
%$$
%$$
\mbox{Normal}\left\{\frac{\sum_{i=1}^{n_2} W_{2i} \mathunderaccent\tilde{x}_{2i}^\prime}{\sum_{i=1}^{n_2} W_{2i}}
\mathunderaccent\tilde{\beta}, \frac{\sigma^2}{\sum_{i=1}^{n_2} W_{2i}}\right\},
$$
where $W_{21},\ldots,W_{2n_2}$  are the original survey weights in the ps. We have used this same technique
for all the five scenarios (models); in Scenario G below, we assume simple random sampling (i.e., survey weights are equal). Unfortunately, it is difficult to take the variabilities (not quasi randomization) of the mean and variance into consideration, and further study is needed in a super-population model-based analysis like what we attempt here. 
One way to do so is to bootstrap the probability sample; see details in Appendix D.

\begin{center}
{\bf 3.3 Additional Observations}
\end{center}

We make three important observations to conclude this section. The first observation is about selection bias,
the second observation is about discounting, the third observation is about the
robustness of posterior inference about the finite population mean to non-normality.
Note that both the ps model and the nps model are adjusted population models, adjusted by the ps weights and nps weights respectively and possibly the discounting factor. The nps weights are estimated
based on a participation model (see CLW) and ps weights are known. 

%First, we discuss selection bias. Without survey weights, both the nps and ps
%are corrupted.  However, if the population 
%model is correct, there is no issue with the ps because the survey weights are known, and they are incorporated 
%into the ps model; the participation model has nothing to do with this. So let us put the ps aside in this 
%discussion on data integration, and focus on the nps. Let us consider two
%cases, where the participation model is correctly specified and where it is incorrectly specified. In the first
%case, the weights, obtained from the propensity scores will be correct. If the population model is incorrect, the selection bias cannot be corrected; if the population model is correct, the selection bias will be corrected. On hand, if
%the participation model is incorrect and it is mis-specified (e.g., the last covariate is deleted), then
%survey weights will be incorrect, and we cannot account for the selection bias even if the population model
%is correct. Therefore, both the participation model and the population must be correct in the Bayesian 
%approach in this paper. That is, we need both models to be robust against their assumptions, and the
%issue of double robustness (see CLW) is not appropriate in our work. 

First, we discuss selection bias. We consider the nonignorable selection model, 
$$
f(R,y \mid \uwd{z}) = f(y \mid R, \uwd{z}) P(R \mid \uwd{z}) = P(R \mid y,\uwd{z})f(y \mid \uwd{z}),
$$
where $\uwd{z}$ is vector of common covariates in the participation model and $R=1$ if a unit is selected and $R=0$  otherwise. See, for example, Nandram and Choi (2010) 
and  references therein for the standard nonignorable selection model; see also Nandram (2022) for examples
on nonprobability sampling. The first equation is the
pattern-mixture model, and CLW used $P(R \mid \uwd{z})$ to estimate the propensity scores, and the second
equation is the selection model, not used in our work here, but this can also be used because we have
the study variable from both the nps and the ps.

Clearly, 
$f(y\mid \uwd{z})$, is obtained by summing  over the two values of $R$, and if $f(y\mid \uwd{x})$
is used to fit the nps data, there is likely to be selection bias unless $f(y\mid \uwd{x})$ is adjusted.
%In our context, this model is not usable because we do not observe  the nonsamples ($\uwd{x}$ and $y$).
 Note that if $\uwd{z} = \uwd{x}$ and $R$ and $y$ are independent given $\uwd{x}$, 
(i.e., $f(y \mid R, \uwd{x}) = f(y \mid \uwd{x}$)), there will be no selection bias,
and one can make inference about the finite population mean without survey weights.
We assume that the population model, $f(y\mid \uwd{x})$, is always
correct, for every unit in the population.
However, the available covariates from the ps may be correlated with hidden covariates, 
and in this case the survey weights may contain additional information. 
It is true that the survey weights from the ps might contain information beyond the observed covariates. 
The population model will be biased, if it is not 
adjusted to include the survey weights for the nps and the ps. 
Another case is when the same covariates are used in the population model and the participation model
(i.e., $\uwd{x} = \uwd{z}$), there can still be selection bias. For example, all the covariates are
important in the participation model and at least one of them is not important in the population model.
In this case, for the nps the population model must also be adjusted using the survey weights. 

Second, we discuss discounting. If the nps is used to construct the prior of the
parameters in our approach and the ps to construct the likelihood of the parameters, the nps 
will dominate it because of its size. 
%the estimated variance will be too small because the nps is large (Big Data, Meng 2018) 
%relative to the ps, and so it will dominate the ps. 
Within the Bayesian paradigm, this is a bad strategy to 
have the prior to dominate the likelihood.  This is why we need the discounting factor in the nps when 
it is used to construct the prior of the parameters, but clearly we do not need discounting when the nps is used
to construct the likelihood of the parameters;
we do need to estimate the survey weights. When the ps is used to construct the prior of the parameters, 
we need to discount it,
because it is really a prior based on past data. But such discounting is expected to be small 
because the nps is relatively much larger; see Appendix E for more discussion.  Then, whether the ps or the nps is used for prior construction,
the prior for the parameters comes from
$$
\pi(y \mid a, \uwd{x}) = \frac{(f(y\mid \uwd{x}))^{aw}}{\int (f(y\mid \uwd{x}))^{aw} dy},  0<a\leq 1.
$$
Given $a$ and $\uwd{x}$, this is a proper distribution in $y$.
It is worth noting  that it is actually the posterior density
of the parameters from the historical data (nps or ps) that is being penalized.

A secondary use of the survey weights is that they serve to 
increase the flexibility of the sample model because it now has heterogeneous variances.
We always used the normalized density with adjusted weight, $w$,
$$
g(y \mid \uwd{x}) = \frac{(f(y\mid \uwd{x}))^{w}}{\int (f(y\mid \uwd{x}))^w dy},
$$
a proper probability density function, to model the sample data only. This is the same for the 
nps and the ps except that in the nps the survey weights must be estimated, and assumed known;
CLW estimated  $P(R \mid \uwd{x})$ for the samples using a weighted likelihood to
account for the selection bias. 
Surrogate sampling is then used to perform
predictive inference by sampling the entire population via the population model.
The normalization constant, $\{\int (f(y\mid \uwd{x}))^w dy\}^{-1}$,
is necessary if it is a function of the other parameters because if it is eliminated, we 
cannot make proper Bayesian analysis (e.g., standard Bayesian diagnostics are not available).

Third, we argue here that the posterior inference about the finite population mean is approximately robust
against non-normality. We have $\bar{Y} \mid \uwd{\beta}, \sigma^2 \sim \mbox{Normal}(\bar{X}^\prime \uwd{\beta}, 
\frac{\sigma^2}{N})$. By the Central Limit Theorem, this is also true approximately without normality. 
Using the results in Appendix C, it is easy to show that
$$
\frac{\bar{Y}-\bar{X}^\prime \hat{\uwd{\beta}}} {\sqrt{(\frac{1}{N} + d\bar{X}^\prime A^{-1} \bar{X})/(n_1+n_2-p)}}
\mid a, \uwd{y} \sim t_{n_1+n_2-p},
$$
where $\hat{\uwd{\beta}}$, $A$ and $d$ are functions of $(w_{si},\uwd{x}_{si},y_{si}),i=1,\ldots,n_s, s=1,2$. Because  $n_1>>n_2$ and
the population size, $N$, is very large, we have 
$$
\frac{\bar{Y}-\bar{X}^\prime \hat{\uwd{\beta}}} {\sqrt{d\bar{X}^\prime A^{-1} \bar{X}/(n_1+n_2-p)}}
\mid a, \uwd{y} \stackrel{approx} \sim \mbox{Normal}(0, 1),
$$
regardless of the distribution of $y_1,\ldots,y_N \mid \uwd{\beta},\sigma^2,a$. Therefore, posterior inference
about $\bar{Y}$ is approximately robust against non-normality (i.e., the normality assumption in the population model is approximately irrelevant).
This is true for Models B, C and D with the discounting factor in B being $a=1$; see Section 4. It is important
to note that the posterior mean and variance are functions of $(w_{si},\uwd{x}_{si},y_{si}),i=1,\ldots,n_s, s=1,2$. While the $w_{1i}$ are
based on estimated propensity scores and posterior inference about $\bar{Y}$ is approximately robust against non-normality, there is no guarantee that it is robust against departures from the participation model that is very important in our work.

\begin{center}
{\bf 4. Numerical Analysis on BMI Data and Related Issues}
\end{center}

In this section, we consider numerical analysis of the BMI data. 
Specifically, in Section 4.1, we use the example on BMI data to compare different scenarios (models), 
and in Section 4.2, we present a simulation study.

As a preliminary analysis, we have looked at the probability sample only, $(W_{2i},\uwd{x}_{2i},y_{2i}),i=1,\ldots,n_2$.
By simply bootstrapping these data, we can provide distributions for $N, \bar{\uwd{X}}, \bar{Y}$. We selected
$B = 10,000$
Bayesian bootstrap (Rubin 1981) samples, each represented by $(W^\ast_{2i},x^\ast_{2i},y^\ast_{2i}),i=1,\ldots,n_2$.
Then, we computed $N_b = \sum_{i=1}^{n_2} W^\ast_{2i}$, $\bar{\uwd{X}}_b = \frac{\sum_{i=1}^{n_2} W^\ast_{2i} 
x^\ast_{2i}}{N_b}$ and $\bar{Y}_b = \frac{\sum_{i=1}^{n_2} W^\ast_{2i} y^\ast_{2i}}{N_b}, b=1,\ldots,B$. 
Let $T^{(1)},\ldots,T^{(B)}$ (ordered smallest to largest) denote the $B$ bootstrap samples of quantity $T$, a 95\% credible interval for $T$
is $(T^{(.025B)}, T^{(.975B)})$, the posterior mean (PM) is the sample average and the posterior standard deviation
is the sample standard deviation. For the BMI data, we have
a 95\% credible interval for
$\bar{Y}$ is $(25.008, 27.115)$ with posterior mean, $PM = 26.002$ and posterior standard deviation, $PSD = .534$. 
%Note that the posterior variance of $\bar{Y}$ is inversely proportional to $N \approx 2,000,000$, thereby making the 
%PSD very small.  
The key question is, ``Can we keep PM the same and considerably reduce the PSD?'' Also, a 95\% credible interval for 
$N$ is $(1,946,029, ~2,837,823)$ and
for the elements of $\bar{\uwd{X}}$, they are respectively $(42.791,~50.380)$, $(.012,~.060)$ and $(.437,~27.115)$, corresponding to age, race
and sex. It is possible to use the Bayesian bootstrap distributions to express uncertainty in the prediction about the inverse probability weighted estimators. To account for variability of $N$, $W_{1i}$, $\bar{X}$, this Bayesian bootstrap procedure can be coupled with the Bayesian method; see Section 4.2.

We describe five scenarios (models), which we denote by  B, C, D, E, G. We want to see how the
weights and the discount factor change the Bayesian predictive inference. The five scenarios (models) are given next.
\begin{itemize}
%   \item[i.] A uses the nps to make inference; the weights from the ps are donated by matching the covariates.
   \item[i.]  B uses only (see Section 3) the nps to make inference; the weights are obtained via propensity scores assisted by the ps.
   \item[ii.] C uses both the ps and the nps; the nps is used as the prior and there is partial discounting (i.e., $0<a<1$).
	 \item[iii.]  D uses both the ps and the nps; the ps is used as the prior and there is partial discounting. 
	 \item[iv.] E uses only the ps  with the survey weights.
%	 \item[vi.] F uses only the nps; data fusion is not used to obtain the weights, and therefore, it is different from A and B.
	 \item[v.] G uses only the ps but weights are omitted, and therefore G is different from E. However, $N$ and $\bar{X}$ are still unknown, and we have used the estimated $N$ and $\bar{X}$ from the ps.
\end{itemize} 
A comparison of these scenarios is informative, and it provides important clues on survey weights, discounting and
data integration.

\begin{center}
{\bf 4.1 BMI Data}
\end{center}

We analyze the BMI data on the 8 counties in California as a nonprobability sample and
a probability sample, otherwise there are no distinctions among the counties.
We note that the correlation between the BMI 
values and the survey weights for the probability sample is 
small ($\approx -.145$), but there are still important differences among the five scenarios. 

\begin{table}[htb]
\begin{small}
%\begin{scriptsize}
\caption{Comparison of five models using BMI data}
\label{Tab:data1}
% see Tab:con1.
\begin{center}
\begin{tabular}{ccccccccccccccccccccccccccc}
%\\
\hline
\\   
   Model &&       PM      && PSD   &&  PCV &&     95\% CI\\
		\hline
\\
    B   && 27.321 &&  0.153 &&    0.006 &&  (27.029,  27.630)\\
		\\
    C  && 27.045 &&  0.134 &&   0.005 &&  (26.787,  27.301)\\
		\\
    D     && 27.097 &&  0.135 &&  0.005 &&    (26.825,  27.350)\\
		\\
    E   && 25.979 &&  0.299 &&    0.012 &&   (25.395,  26.562)\\
		\\
    G   && 26.856 &&  0.304 &&   0.011 &&   (26.285,  27.470)\\
		
\\
\hline
\end{tabular}
%\\
%\vspace*{.10in}
\flushleft
NOTE:  The models are B: nps only; C: data integration with nps as prior; D: data integration with ps as prior; E: ps only;
G: ps without survey weights. When the nps is used as the prior,
the 95\% HPD interval for the discount factor ($a$) is $(0.670, 0.945)$
and when ps is used as the prior, it is $(0.989, 0.999)$.
\end{center}
%\end{scriptsize}
\end{small}
\end{table}
 
In Table \ref{Tab:data1}, we present posterior summaries of the finite population mean.
We use  the posterior mean (PM), posterior standard deviation (PSD), numerical 
standard error (NSE, not really necessary here), posterior coefficient of variation (PCV)
and 95\% credible interval. 
%We see that the PM of E is different from the other four models that are very similar.
As expected, there are two groups: E and G (no data integration) and B, C and D (data integration)
with the respective group members being similar in terms of PM and PSD.
For PSDs, B, C, D are small and those for E and G are much larger. NSEs and PCVs are very good for all
models. The 95\% HPD intervals for B, C, D are similar and to the right of those of E and G, which 
have wider intervals. As expected, E should be unbiased with large PSD.

The discount factor $a$ is significant when the nps is used as the prior [95\% HPD interval is $(.670,.945)$]
and it is nearly $1$ when the ps is used as the prior [95\% HPD interval is $(.989,.999)$]. This is sensible 
because the ps is small (not much to discount; see Appendix F for an illustration) and of high quality and 
the nps is large but of low quality.
%
%\begin{figure}[h!]
  %\centering
%%  \includegraphics[width=.9\textwidth, height=.9\textheight]{fpmplot-new-BCDEG.ps}
	  %\includegraphics[width=.7\textwidth, height=.7\textheight]{fpmplot-new-BCDEG.pdf}
  %\caption{\label{Fig:dat} Comparison of the posterior distributions of the finite population mean for the five models 
	%(B, C, D, E, G)}
%\end{figure}

\begin{figure}[h!]
    \centering
	  \includegraphics[width=.7\textwidth, height=.7\textheight]{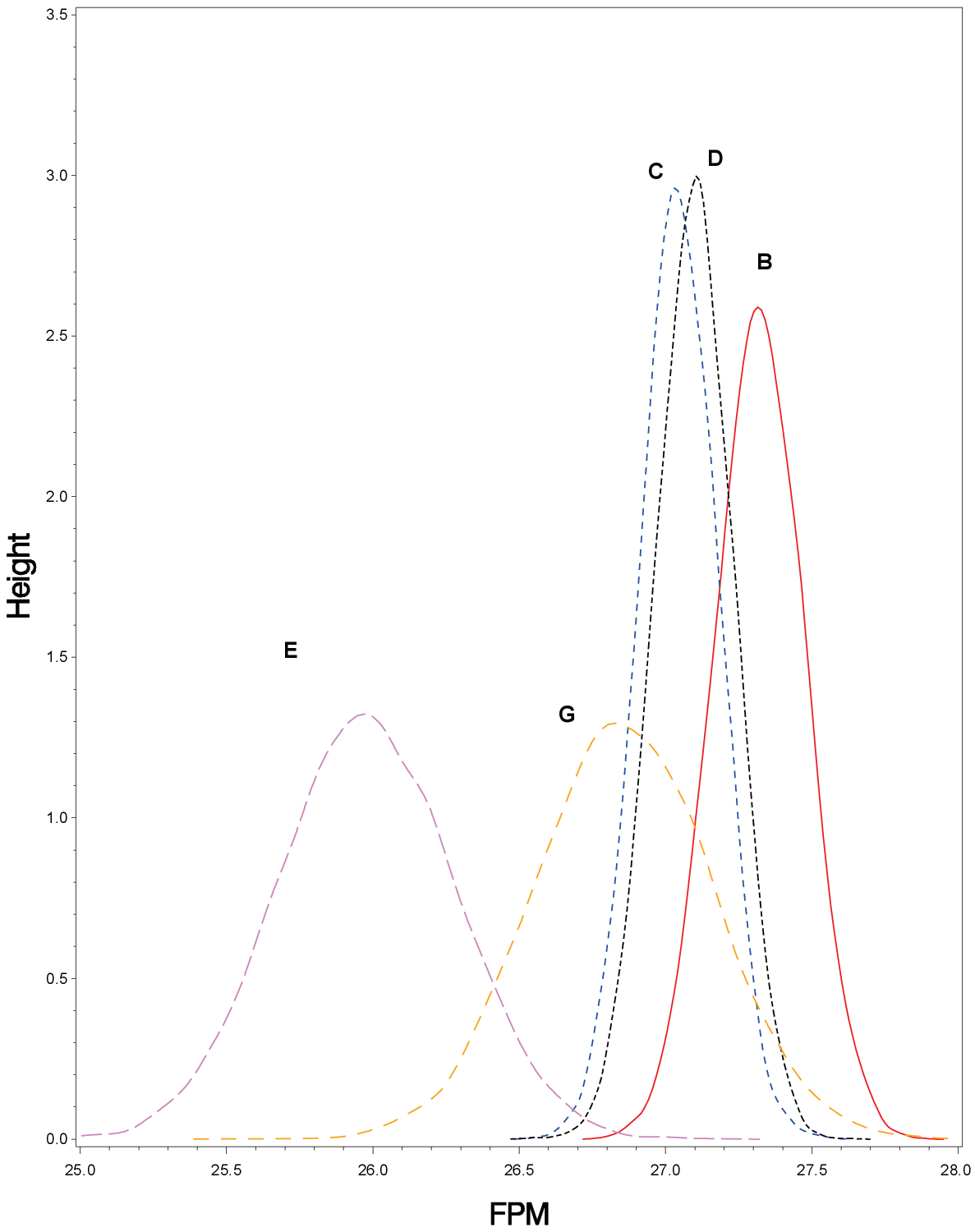}
  \caption{Comparison of the posterior distributions of the finite population mean for the five models 
	  (B, C, D, E, G)}		   
    \label{Fig:dat}
\end{figure}

%
%\begin{figure}[h]
    %\centering
    %\includegraphics{filename.ps}
    %\caption{Caption text}
    %\label{fig:labelname}
%\end{figure}

%
%\begin{figure}[h!]
  %\centering
  %\includegraphics[width=.9\textwidth, height=.9\textheight]{fpmplotB.pdf}
  %\caption{\label{Fig:dat} Comparison for the posterior distributions of the finite population mean by the five models 
	%(B, C, D, E, G)}
%\end{figure}

In Figure \ref{Fig:dat}, we have compared the posterior densities of the finite population mean for the five methods.
All posterior densities are unimodal with B, C and D concentrated around 27; E and G are different from
these. In the plot, E is to the left; E and G have very large spread.

We have looked at similar posterior summaries for the regression coefficients and the 
variance, $\sigma^2$. We note that SWRB and WSRB were mainly concerned about regression coefficients.
Inference about $\beta_1$, the intercept, is similar over the five models. However,
there are some differences for the other regression coefficients. For models C and D, all parameters are important. In models A, E and G, $\beta_3, \beta_4$ are important; for models B and G, $\beta_2$
is also important. The PSDs for models B, C, D and E, which are similar (D has the smallest PSD), are much smaller than those for model G. 

Finally, in Appendix D, we show how to incorporate uncertainty about the propensity scores
in the sample model for the illustrated example on BMI.

%\begin{figure}[h!]
  %\centering
%%  \includegraphics[width=.9\textwidth, height=.9\textheight]{fpmplot-new-BCDEG.ps}
	  %\includegraphics[width=.7\textwidth, height=.7\textheight]{fpmplot-new-BCDEG.ps}
  %\caption{\label{Fig:dat} Comparison for the posterior distributions of the finite population mean by the five models 
	%(B, C, D, E, G)}
%\end{figure}

%%%%%%%%%%%%%%%%%%%%%%%%%%%%

\begin{center}
{\bf 4.2 Simulation Study}
\end{center}

%We follow CLW design-based method for generating the finite population and the
%sample. 
We report the results of a simulation study following the CLW design-based method for generating
the finite population and the samples.
However, we associate the data simulation within the framework of the BMI data. The BMI data 
have the following structure,
$$  
E(y_i) = 23.8449 + .0559 x_{1i} + 2.2656 x_{2i} + .0452 x_{3i}, i=1,\ldots,N
$$
at least for the samples (nps and ps). We adjust this a bit by replacing $.0452$ by
$.2525$ to avoid computational instability.
Following CLW, we have taken $N=20000$, $n_1=1500$ and $n_2=300$. 
%We use no other values.

Modifying the sampling process of CLW, we perform the following steps, for $i=1,\ldots,N$,
\begin{itemize}
  \item[i.] Draw 
$$
x_{1i} \stackrel{ind} \sim \mbox{Uniform}(20, 90)
$$
and set $b_i = \{23.8449 +.0559 x_{1i}\}^\frac{1}{10}$;

\item[ii.] Draw 
$$
x_{2i} \mid x_{1i} \stackrel{ind} \sim \mbox{Bernoulli}\{\frac{e^{b_i}}{1+e^{b_i}}\}
$$
and set $b_i = \{23.8449 +.0559 x_{1i} + 2.2656x_{2i}\}^\frac{1}{10}$;

\item[iii.] Draw
$$
x_{3i} \mid x_{1i} x_{2i} \stackrel{ind} \sim \mbox{Bernoulli}\{\frac{e^{b_i}}{1+e^{b_i}}\};
$$

\item[iv.] Finally, construct
$$
y_i = 23.8449 + .0559x_{1i} + 2.2656 x_{2i} + .2525 x_{3i} +  e_i, e_i \stackrel{iid} \sim \mbox{Normal}(0, \sigma^2).
$$
\end{itemize}
This gives us the finite population of values $(\uwd{x}_i,y_i),i=1,\ldots,N$. So we can compute the true value
of $\bar{Y} = \frac{1}{N} \sum_{i=1}^N y_i$. Following CLW, we have selected $\sigma^2$ such that
the correlation, $\mbox{Cor}(23.8449 + .0559x_{1i} + 2.2656 x_{2i} + .2525 x_{3i}, y_i) = \rho$,
and we have selected $\rho$ as $\rho = ..20,  .30, 50, .80$. This is done by trial and error.

\begin{table}[htb]
%\begin{scriptsize}
\begin{small}
\caption{Simulation Study: No Misspecification}
\label{Tab:sima}
\begin{center}
\begin{tabular}{ccccccccccccccccccccccccccc}
%\\
\hline
\\
%\multicolumn{10} {c} {\underline{Model}}  \\
&&&&&& Model &&&&\\
\hline
\\
 Measure  && $\rho$ &&  $B$  &   $C$    &  $D$  &    $E$   &   $G$\\
\\
\hline
\\
ARB &&  0.20  &&  0.006  &   0.006  &   0.005  &   0.017  &   0.015\\
    &&  0.30  &&  0.003  &   0.003  &   0.003  &   0.010  &   0.008\\
    &&  0.50  &&  0.002  &   0.002  &   0.002  &   0.006  &   0.005\\
    &&  0.80  &&  0.001  &   0.001  &   0.001  &   0.002  &   0.002\\
\\
PRMSE &&  0.20  &&  0.282  &   0.270  &   0.264  &   0.705  &   0.730\\
    &&  0.30  &&  0.164  &   0.160  &   0.155  &   0.421  &   0.424\\
    &&  0.50  &&  0.091  &   0.089  &   0.086  &   0.234  &   0.236\\
    &&  0.80  &&  0.039  &   0.038  &   0.037  &   0.101  &   0.101\\
\\
Cov && 0.20 &&   0.997   &  0.944   &  0.978   &  0.823   &  0.939\\
    && 0.30 &&   0.999   &  0.949   &  0.980   &  0.834   &  0.960\\  
	  && 0.50 &&   0.999   &  0.948   &  0.980   &  0.833   &  0.960\\
    &&  0.80 &&   0.999  &   0.949  &   0.980  &   0.834  &   0.960\\
\\
Wid && 0.20 &&   0.863   &  0.757   &  0.771   &  1.698   &  2.064\\
    && 0.30 &&   0.526   &  0.461   &  0.471   &  1.034   &  1.254\\
    && 0.50 &&   0.292   &  0.257   &  0.262   &  0.575   &  0.698\\
    && 0.80 &&   0.126   &  0.110   &  0.113   &  0.247   &  0.300\\
\\
\hline
\\
\end{tabular}
\flushleft
%NOTE:   Here,
%$x_3$ is used in both the response model and the participation model.
\end{center}
%\end{scriptsize}
\end{small}
\end{table}

The selection probabilities for nps are
$$
\pi_{1i} = \exp(b_i)/(1+\exp(b_i), b_i= \theta_0 + .1x_{1i} + 0.2x_{2i} + 0.1x_{3i}, i=1,\ldots,N,
$$
and $\theta_0$ is selected by trial and error such that $\sum_{i=1}^N \pi_{1i} = n_1$.
We select $\pi_{2i}$ such that
$$
\pi_{2i} = n_2 z_i/\sum_{i=1}^N z_i, z_i = \theta_1 + x_{1i} + 0.2x_{2i} + 0.1x_{3i},
$$
where $\theta_1$ is selected, again by trial and error, to ensure $\max\{z_i\}/\min\{z_i\} \approx 50$.
This deviates a little bit from CLW.
As in CLW, the nps is taken using Poisson sampling with probabilities $\pi_{1i}$ and target sample size $n_1$,
and the ps taken using randomized systematic PPS sampling with target sample size $n_2$. 
%
%Finally, we get the true values of the regression coefficients using the least squares estimates,
%$$
%y_i \stackrel{ind} \sim (\uwd{x}_i^\prime \uwd{\beta}, \sigma^2), i=1,\ldots,N,
%$$
%$\hat{\uwd{\beta}} = (X^\prime X)^{-1} X^\prime \uwd{y}$, where $X$ is the design matrix. Of course, $\hat{\uwd{\beta}}$ is 
%based on the population model and because $N$, $N=20000$, is large, $\hat{\uwd{\beta}}$ is almost a point mass. 
%CLW did not study the regression coefficients.

We have run the simulations as follows.
\begin{itemize}
 \item[a.]  For each setting of $\rho$, we have generated one finite population, and we took 1000 samples (a nps and a ps) from it. We use the following notations:
$T$ is true finite population mean, $PM$ is the  posterior mean, $PSD$ is posterior standard deviation and
 $(C025, C975)$ is the 95\% highest posterior density interval (HPDI);
\item[b.]  We computed the absolute relative bias: $ARB = \mid (PM-T)/T \mid$; posterior root mean squared error, $PRMSE = \sqrt{(PM-T)^2 + PSD^2}$ and
incidence: $I = 1$ if a 95\% HPDIs containing $T$, $0$ otherwise; width, $Wid = C975-C025$;
\item[c.] Finally, we averaged the $1000$ runs; coverage, $Cov$, is the proportion of HPDIs containing $T$.
\end{itemize}

In Table \ref{Tab:sima}, we present simulation comparisons. 
The smallest ARBs come from B, C, D, and E, G have slightly larger ARBs. The PRMSEs of B, C, D are smaller than E, G with C, D
slightly smaller than B. The coverage for E is below the nominal value of 95\%; B and D are too conservative but C is
just about the nominal value. For Wid, C and D dominate the others, considerably shorter than B, and E and G are too
wide. We note that ARB, PRMSE and Wid decrease with increasing $\rho$; it is clear that C is 
the winner and B and D are competitive. 
For  the discount factor $a$, the $PM  ~(PSD)$, averaged over the simulation runs, are  for C $.569 ~(.051)$ (i.e., considerable discounting) and 
for D  $.981 ~(.016)$ (i.e., no discounting) with very little changes  over $\rho$. This indicates
that one should use the nps as the prior with a penalty and inference should be made using the ps (poor coverage although very
wide) integrated with the nps. We have seen similar results in the example on BMI data.

We have also looked at two cases of mis-specifications; see the discussion at the end of Section 3. In both cases, for the nps
the third covariate is used to estimate the propensity  scores in the participation model. In the first case, the third covariate is omitted from the population (regression) model after it is obtained in the simulated study variable. In the second case, 
the third covariate is not used in the data simulation of the study variable and must be omitted from the population model. 
%the third covariate is used in getting the propensity  scores in the participation model. 
We found that the two 
models with discounting are competitive with  the others, but for all models coverage decreases as 
$\rho$ increases. This must be true because larger $\rho$ means smaller $\sigma^2$ in the simulation runs, thereby making the coverage smaller as $\rho$ increases. 
Finally, we note that one drawback of our procedure is that 
both the population model and the participation model must be correctly specified, and so robustness
is essential. We will address this issue.
%(a possible defect in the simulation design of CLW). 

%Perhaps this needs to be looked at more carefully.

%%%%%%%%%%%%%%%%%%%%%%%%%%%%%%%%%%%%%%%%%%%%%%%%%%%%%%%%%%%%%%%%%%%%%%%%%%%%%%%%%%%%%%
 
\begin{center}
{\bf 5. Concluding Remarks}
\end{center}

Two important findings show up in our work. First, if one has the study variable on both the ps and the nps, 
and the nps is much larger than the ps, as is usually the case, then it is better to use the nps to construct the prior with partial discounting. Second, if the ps is used to construct the prior with partial discounting, there 
will be virtually no discounting; see Appendix E for an illustration. Apparently, this is sensible because the ps is much smaller than the nps and of higher quality, although we assume no measurement errors. It is erroneous to use the ps as the prior or to use the ps to supplement the nps, rather one should use the nps to supplement the ps, provided the study variable is available in the ps. This concurs with Sakshaug et al. (2019) and others.
In Appendix E, we further discuss the importance of the ps as a prior.

Although we have assumed normality on the study variable, our data integration methodology is very general.
We can use any reasonable distribution for the study variable (e.g., skew normal). We can also make inference
about other finite population parameters such as finite population quantiles (e.g., $95^{th}$ percentile for BMI,
a measure of obesity); albeit with more computational effort to sample the entire population. In Appendix A, 
we show how our method works when the study variable is binary. 

%Extensions to finite population
%quantiles are under investigation.
%We note that it is not possible to estimate 
%finite population quantiles within the framework of Chen, Li and Wu (2020) nor can they use the study variable
%in the ps, but we can.

There is a need to express uncertainty in the estimation of propensity scores in the Bayesian approach;
it is important to account for the variability in the propensity scores.
%because  there is no such consideration in
%the method of Chen, Li and Wu (2020). Recall that the MLEs under the logistic regression model are obtained from the %quasi-likelihood as in Chen, Li and Wu (2020) and assumed known; see Appendix A. 
In Appendix D, we have shown how to use the bootstrap (Bayesian or non-Bayesian) method to incorporate
uncertainty in the estimated survey weights. There is considerable underestimation in Scenario B with C and D showing much less underestimation; there is less underestimation in Scenario E because the ps weights are known.  However, the
bootstrap is not the best way to do this; one would need a model to contain the unknown survey weights with estimation
being done in the same model. Within the Bayesian paradigm, this is a difficult problem and it is under study.
While we can link the parameters of the participation model to the adjusted survey weights in the sample model,
it turns that the computation in the unified model is difficult.

Also, it is possible to have the discounting factors vary with the observations. For example, using the nps as the prior,
we can replace $a$ by $a^{1/w_{1i}}$, reflecting less discounting for observations with larger survey weights.
We have not attempted this problem yet.

The assumption of normality on the BMI data is perhaps not a very good one because the BMI data are skewed and discrete;
see Yin and Nandram (2020 a,b). 
Also, more robust methods on propensity scores are needed. 
%The generalized Dirichlet
%process with general stick-breaking priors (e.g., Pitman-Yor process) can be used to provide more robust models, but these models  are difficult to fit  when all uncertainty is taken into account; see Ishwaran and James (2001). 
It is also possible
to use BART in data integration (e.g., Rafei, et al. 2021); one does not need to express a relation between study variable
and covariates. But BART is not a fully Bayesian procedure because it double-uses the data, it suffers from overshrinkage, and there is no underlying theory of BART; see Hill, Linero and Murray (2020) for further details.
%It is possible to relax the parametric assumption through infinite mixtures and this will be studied in a separate paper.

More importantly, for the study variable a more robust population model is needed. 
A stick-breaking prior with finite mixture and the Pitman-Yor process can be used to provide
more robust population models; see Ishwaran and James (2001). This will be studied in a separate paper.
%Through infinite mixtures, the generalized Dirichlet process with general stick-breaking priors (e.g., Pitman-Yor %process) can be used to provide more robust population models; see Ishwaran and James (2001). This will be studied in %a separate paper.
%Appendix E gives more detail.
%%%%%%%%%%%%%%%%%%%%%%%%%%%%%%%%%%

\newpage
\vspace{.10in}
{\bf Acknowledgments}
\vspace{.25in}

%Balgobin Nandram thanks Jai Won Choi for discussions and his students,  Ashley Lockwood,
%Lingli Yang and Yang Liu, for providing corrections to an earlier manuscript. 
Balgobin Nandram
gave invited presentations on different versions of this paper at the 2021 annual meeting of Statistical Society of Canada,
the 2021 Joint Meetings of the American Statistical Association (both virtual), the 2022 SAE annual meeting at the University of Maryland (in person), and Banaras Hindu University, India (in person) in 2023. Balgobin Nandram  was supported by a grant from the Simons Foundation
(\#353953, Balgobin Nandram) and J. N. K. Rao was supported by a research grant from the Natural Sciences and Engineering Research Council of Canada.

%change the sentence to "Nandram's work ...Nandram) and J. N. K. Rao was supported by a research grant from the %Natural Sciences and Engineering Research Council of Canada.
%%%%%%%%%%%%%%%%%%%%%%%%%%%%%%%%%%%%%%%%%%%%%%%%%%%%%%%%%%%%%%%%%%%%%%%%%%%%%%%%%%%%%%

\vspace{.25in}

\begin{center}
{\bf APPENDIX A: Model for Binary Study Variables} 
\end{center}
\def\theequation{A.\arabic{equation}}
\setcounter{equation}{0}

We consider the case in which the study variable, $y$ is binary, and with covariates we use  logistic
regression. Let $(W_{si}, \uwd{x}_{si}, y_{si}), i=1,\ldots,n_s, s=1,2$, where $s=1$ refers to the nps
and $s=2$ to the ps. We have estimated $W_{1i}$ using the CLW method, and we assume they are known as
in the normal case. As before, we denote the adjusted weights by $w_{si}, i=1,\ldots,n_s, s=1,2$. Note
little $w_{si}$ and big $W_{si}$. Let $a_1 = a$ and $a_2=1$ denote the discounting factors (i.e., no
discounting for ps). 

The population model is
\begin{equation}
\label{bpop}
P(y_{i} = 1 \mid \uwd{\beta}) = \frac{e^{\uwd{x}_i^\prime \uwd{\beta}}}
{1+e^{\uwd{x}_i^\prime \uwd{\beta}}}, i=1,\ldots,N,
\end{equation}
where the population size, $N$, may be unknown and the nonsampled covariates, $\uwd{x}_i, i=n+1,\ldots,N$,
are also unknown.

As in the normal case, for the sample the probability mass function of $y_{si}$ is obtained by
normalization,
$$
P(y_{si} = 1 \mid a, \uwd{\beta}) = \frac{e^{a_sw_{si}\uwd{x}_{si}^\prime \uwd{\beta}}}
{1+e^{a_sw_{si}\uwd{x}_{si}^\prime \uwd{\beta}}},  i=1,\ldots,n_s, s=1,2
$$
and there is independence over $i$.
Then, using the noninformative prior,
$$
\pi(a, \uwd{\beta}) = 1, 0 \leq a \leq 1, \uwd{\beta} \in R^p,
$$
where we assume $p-1$ covariates plus and intercept. Letting $\uwd{y} = (\uwd{y}_{s}, s=1,2)$, the joint posterior density is
\begin{equation}
\label{bp}
\pi(a, \uwd{\beta} \mid \uwd{y}) = \prod_{s=1}^2 \prod_{i=1}^{n_s} \left \{\frac{e^{a_sw_{si}\uwd{x}_{si}^\prime \uwd{\beta}y_{si}}}
{1+e^{a_sw_{si}\uwd{x}_{si}^\prime \uwd{\beta}}} \right \}.
\end{equation}
It can be shown that if $X_2 = (\uwd{x}_{2i}^\prime)$ is full rank, $\pi(a,\uwd{\beta} \mid \uwd{y})$ in (\ref{bp}) is
proper. Samples from $a, \uwd{\beta} \mid \uwd{y}_s,s=1,2$ can drawn using the griddy Gibbs sampler. Denote the sample of
size $M$ from the Gibbs sampler as $(a^{(h)},\uwd{\beta}^{(h)}), h=1,\ldots,M$.
[A Gibbs sampler is not needed in the normal case.] 

However, in the case of binary study variables, prediction is more difficult and time-consuming than in the
normal case. The main difficulty is that one would need to sample all the population covariates (assumed unknown) subject to the constraint, 
\begin{equation}
\label{con}
\sum_{i=1}^N \uwd{x}_{i} = \sum_{i=1}^{n_2} W_{2i}\uwd{x}_{2i},
\end{equation}
where we estimate the 
population size, $N$,  by $\sum_{i=1}^{n_2} W_{2i}$. We can resample $\uwd{x}_{si}, i=1,\ldots,n_s, s=1,2$,
until the constraint is met using rejection sampling with some tolerance.  For each $(a^{(h)},\uwd{\beta}^{(h)})$ from the Gibbs
sampler, we will perform the procedure to get all $\uwd{x}_i^{(h)},i=1,\ldots,N$, denote by $\uwd{x}^{(h)}$. Therefore, we now have
$(a^{(h)},\uwd{\beta}^{(h)}, \uwd{x}^{(h)}), h=1,\ldots,M$, which will be use to get $\bar{Y}^{(h)}, h=1,\ldots,M$.
It is possible to operationalize 
the sampling procedure of the population covariates by discretization of the  covariates. In our application, we 
need to discretize age when we study obesity (i.e., binary variable)  as race and sex are binary. 

Finally, we draw surrogate samples from the population model (\ref{bpop}), and compute
the finite population proportion, $\bar{Y} = \frac{1}{N} \sum_{i=1}^N y_i$ at $(\uwd{x}_1,\ldots,\uwd{x}_N)$.
Here, given $\uwd{\beta}$, $y_i=1$ if $u_i \leq e^{\uwd{x}_i^\prime \uwd{\beta}}/(1+e^{\uwd{x}_i^\prime 
\uwd{\beta}})$ and $y_i = 0$ otherwise, with $u_i \stackrel{iid} \sim \mbox{Uniform}(0,1), i=1,\ldots,N$.
To estimate the posterior distribution of $\bar{Y}$,
this is done for each iterate from the Gibbs sampler together with each set of  population covariates resampled 
subject to the constraint (\ref{con}).

\begin{center}
{\bf APPENDIX B: Calibration}
\end{center}
\def\theequation{B.\arabic{equation}}
\setcounter{equation}{0}

We show how to calibrate the nps to the ps. Here,  we simply need the population estimated totals from the ps.
We are assuming that the ps is very small, so that the estimated totals from the ps are not very reliable for calibration. We can obtain the totals from a census, administrative records or web scraping. Let $\uwd{t}$ denote the 
vector of the $p$ totals including the intercept; note that it appears 
Haziza and Beaumont (2017) did not use the intercept but this is necessary. Generally, the basic weighting system
ensures consistency of a survey with a census by reducing nonsampling errors (e.g., response errors and coverage
errors) and improves precision; see Haziza and Beaumont (2017) for more discussion. Deville and Sarndal (1992)
presented a general theory of calibration.
 
Let the original survey weights in the ps be $w_j, j=1,\ldots,n$. [Momentarily we drop the subscript
on $n$.]  We search for a calibrated weighting system
$\tilde{w}_j, j=1,\ldots,n$, such that $\sum_{j=1}^n \tilde{w}_j \uwd{z}_j = \uwd{t}$, the calibration
equations. We want the $\tilde{w}_j, j=1,\ldots,n$, to be as close as possible to $w_j, j=1,\ldots,n$.
Haziza and Beaumont (2017) judged closeness by a distance function, $G(u)$, where
\begin{itemize}
   \item[a.] $G(u) \geq 0$ and $G(1) = 0$;
   \item[b.] $G(u)$ is differentiable, $g(u) = G^\prime(u), g(1) = 0$, and strictly convex.
\end{itemize}
They proposed to minimize $\sum_{j=1}^n \frac{\tilde{w}_j}{q_j} G(\frac{\tilde{{w}_j}}{w_j})$ over $\tilde{w}_j,j=1,\ldots,n$,
where $q_j$ denote the importance of unit $j$,
subject to the constraint $\sum_{j=1}^n \tilde{w}_j \uwd{z}_j = \uwd{t}$.
%
 %see Appendix A for details
%on how to solve the Lagrangian system of equations for  a general $G(u)$ (we have made some changes to
%Haziza and Beaumont, 2017). In Appendix A, the weights are given as
%$$
%\tilde{w}_j = w_j g^{-1}(q_j \uwd{\lambda}^\prime \uwd{x}_j), j=1,\ldots,n,
%$$
%where $g(\cdot)$ is the calibration function and $\uwd{\lambda}$ are Lagrangian multipliers.
%The final step on how to numerically compute $\uwd{\lambda}$ is given in Appendix A for 
%a general $G(u)$.

This is done by considering the function,
%Haziza and Beaumont (2017) considered the function,
$$
\phi(\tilde{w_1},\ldots,\tilde{w_n},\uwd{\lambda}) =
\sum_{j=1}^n  \frac{\tilde{w_j} G(\frac{\tilde{w_j}}{w_j})}{q_j}
-\uwd{\lambda}^\prime(\sum_{j=1}^n \tilde{w}_j \uwd{z}_j - \uwd{t}),
$$
where $\uwd{\lambda} = (\lambda_1,\ldots,\lambda_p)^\prime$ are Lagrangian multipliers.
Differentiating  $\phi(\tilde{w_1},\ldots,\tilde{w_n},\uwd{\lambda})$ with respect to $\tilde{w_j}$,
%they got
leads to
$$
\tilde{w_j} = w_j g^{-1}(q_j \uwd{\lambda}^\prime \uwd{z}_j), j=1,\ldots,n.
$$
%Therefore,
The Lagrangian multipliers are determined from
$$
\sum_{j=1}^n w_j g^{-1}(q_j \uwd{\lambda}^\prime \uwd{z}_j) \uwd{z}_j = \uwd{t}.
$$

Here, our method differs from Haziza and Beaumont (2020); they used the Newton-Ralphson
method. We use the Nelder-Mead to minimize $\sum_{k=1}^p \mid \sum_{j=1}^n w_j g^{-1}(q_j \uwd{\lambda}^\prime \uwd{z}_j)
z_{jk}  -t_k \mid$
over $\uwd{\lambda}$, forcing each component down to zero, to get $\hat{\uwd{\lambda}}$. The weights are then,
$$
\tilde{w}_j = w_j g^{-1}(q_j \hat{\uwd{\lambda}}^\prime \uwd{z}_j), j=1,\ldots,n.
$$ 
 
In our case, we use the simple Euclidean distance function for $G(u)$.
%This gives closed-form answers; apparently this was not recognized by Haziza and Beaumont (2017).
We choose $q_j=1,j=1,\ldots,n$.
The Euclidean distant function, $G(u) = (u-1)^2$, is a legitimate distance function because $G(1) = 0$, $g(u) = 2(u-1)$,
$g(1) = 0$, $g^\prime(u) = 2>0$ and so $G(u)$ is strictly convex. Also $g^{-1}(y) = 1 + \frac{y}{2}$.
In this case, we have
$$
\sum_{j=1}^n {w}_j (1+\frac{\lambda^\prime \uwd{z}_j}{2}) z_{jk} = t_k, k=1,\ldots,p.
$$
That is,
$$
\sum_{k^\prime=1}^p \sum_{j=1}^n \lambda_{k^\prime} z_{jk^\prime} z_{jk} = 2(t_k-\sum_{j=1}^n {w}_jz_{jk}), k=1,\ldots,p,
$$
and
$$
A \uwd{\lambda} = \uwd{b}, ~~A = \left(\sum_{j=1}^n w_j z_{j k^\prime}z_{jk} \right)_{(k^\prime,k)},  
~~\uwd{b} = 2(\uwd{t}-\sum_{j=1}^n {w}_j \uwd{z}_{j}).
$$
Therefore, assuming $A$ is invertible, $\hat{\lambda} = A^{-1} \uwd{b}$ and the calibrated weights are
$\tilde{w_j} = w_j(1+\frac{ \hat{\uwd{\lambda}}^\prime \uwd{z}_j}{2}), j=1,\ldots,n$.
It is worth noting that if we used the ps to get $\uwd{t}$, then $\sum_{j=1}^n w_j \uwd{z}_j = \uwd{t}$
and $\uwd{\lambda}= \uwd{0}$; the weights can be negative though (Deville and Sarndal, 1992).
If $\uwd{b}$ is closed to $\uwd{0}$, there will be little difference using
calibration over the nps estimated weights. 

In our example on eight counties in California, we use data obtained from the US Census Bureau on the internet
 to get the totals of the covariates. 
Note that we are calibrating the  nps estimated weights  to those of the ps and not using the estimated totals 
from the ps. We got a population size of $N=4,035,862$, and the other totals are age $= 36.7 \times N$, 
race $= .719\times N$,  sex $ = .497  \times N$. We got $\hat{\lambda}_1 = .000713$, $\hat{\lambda}_2 =-0.000001$, 
$\hat{\lambda}_3 =  0.002198$, $\hat{\lambda}_4 = -0.000124$. It is not surprising then that calibrated weights 
are barely different from the original weights. It is possible that some calibrated weights can be negative,
and negative weights can be set equal unity, with the final weights adding up to the population size.

\begin{center} 
{\bf APPENDIX C: Joint Posterior Density when PS is the Actual Sample}
\end{center}
\def\theequation{C.\arabic{equation}}
\setcounter{equation}{0}

We obtain a random sampler to draw $\uwd{\beta}, \sigma^2, a \mid \uwd{y}$ under the linear 
regression model and we show that the  joint posterior density is proper.

Letting $\uwd{y} = (\uwd{y}_1, \uwd{y}_2)$, the joint posterior density is
$$
\pi(\uwd{\beta}, \sigma^2, a \mid \uwd{y}) \propto a^{n_1/2} (\frac{1}{\sigma^2})^{\frac{n_1+n_2}{2}+1}
e^{-\frac{1}{2\sigma^2} Q}, 0 \leq a \leq 1,
$$
where $Q = a \sum_{i=1}^{n_1} w_{1i}(y_{1i}-\uwd{x}_{1i}\uwd{\beta})^2 + 
\sum_{i=1}^{n_2} w_{i2}(y_{2i}-\uwd{x}_{2i}\uwd{\beta})^2.$
For convenience, letting $a_1=a, a_2=1$ for the case when the nps is used as the prior.
We obtain a random sampler, not a Gibbs sampler, and show that the joint posterior density
is proper at the same time.

First, let us look at $Q$ and assume the design matrix is full rank at least for the ps. We find the conditional
posterior density of $\uwd{\beta}$, which clearly has a multivariate normal density. We now decide its 
mean and variance using a standard trick by differentiation. First, letting $\Delta(\uwd{\beta}) = Q$,
we have
$$
\Delta^{\prime}(\uwd{\beta}) = 2 \sum_{s=1}^2 \sum_{i=1}^{n_s} a_s w_{si} (y_{si}-\uwd{x}_{si}^\prime \uwd{\beta})
\uwd{x}_{si}
$$
and the Hessian matrix is
$$
\Delta^{\prime \prime}(\uwd{\beta}) = 2 \sum_{s=1}^2 \sum_{i=1}^{n_s} a_s w_{si} \uwd{x}_{si}
\uwd{x}_{si}^\prime.
$$
Now, setting $\Delta^{\prime}(\uwd{\beta}) = \uwd{0}$, we get
$$
\uwd{\hat{\beta}} = A^{-1} \uwd{b}, A = \sum_{s=1}^2 \sum_{i=1}^{n_s} a_s w_{si}\uwd{x}_{si} \uwd{x}_{si}^\prime, 
~\uwd{b} =   \sum_{s=1}^2 \sum_{i=1}^{n_s} a_s w_{si}\uwd{x}_{si} y_{si}.
$$
We require $A$ to be nonsingular, a mild assumption because of the size of the nps.
Therefore,
\begin{equation}
\label{c1}
\uwd{\beta} \mid \sigma^2, a, \uwd{y} \sim \mbox{Normal}(\uwd{\hat{\beta}}, \sigma^2 A^{-1}).
\end{equation}

Second, integrating out $\uwd{\beta}$ from the joint posterior density, we get
$$ 
\pi(\sigma^2, a \mid \uwd{y}) \propto a^{n_1/2} \left(\frac{1}{\sigma^2}\right)^{(n_1+n_2-p)/2+1}
{\mid A \mid^{1/2}}
e^{-\frac{1}{\sigma^2} \sum_{s=1}^2 \sum_{i=1}^{n_s}(y_{si}-\uwd{x}_{si}^\prime \uwd{\hat{\beta}} )^2 }. 
$$
Then, letting $d = \sum_{s=1}^2 \sum_{i=1}^{n_s}a_s w_{si} (y_{si}-\uwd{x}_{si}^\prime \uwd{\hat{\beta}})^2$,
\begin{equation}
\label{c2}
\sigma^2 \mid a, \uwd{y} \sim \mbox{InvGam}\left(\frac{n_1+n_2-p}{2}, \frac{d}{2} \right).
\end{equation}

Finally, integrating out $\sigma^2$, we have
\begin{equation}
\label{c3}
\pi(a \mid \uwd{y}) \propto \frac{a^{n_1/2} \mid A \mid^{-1/2}}{d^{(n_1+n_2-p)/2}}, 0 \leq a \leq 1.
\end{equation}

Because $0 \leq a \leq 1$, all quantities are well defined, and provided the design matrix of the ps
is full rank, the joint posterior density is proper. 

Draws can be made from the joint posterior density
using the multiplication rule of probability, drawing samples from (\ref{c3}), (\ref{c2}) and (\ref{c1})
in that order. Samples can be drawn from $\pi(a \mid \uwd{y})$ in
(\ref{c3}) using the grid method.

It is possible to generalize this algorithm  for more robust population models; see Section 5. For 
example, the BMI data are skewed to the right, and the linear regression model may be questionable.

\vspace*{.25in}

\vspace*{.25in}

\vspace*{.25in}
\begin{center}
{\bf APPENDIX D: Incorporating Uncertainty about Unknown Survey Weights} 
\end{center}
\def\theequation{D.\arabic{equation}}
\setcounter{equation}{0}

We use the Bayesian bootstrap to assist in taking care of underestimation of variability.
First, we look at the estimated survey weights for the non-probability sample. Specifically, we incorporate the variability of the estimated weights, $W_{1i}, i=1,\ldots,n_1$, in the models.
Second,  we also assess the variability in estimating the population size, $N$, by $\hat{N} =\sum_{i=1}^{n_2} W_{2i}$ and the population mean covariate, $\uwd{\bar{X}} = \frac{\sum_{i=1}^{N} \uwd{x}_{i}}{N}$, by
$\uwd{\bar{x}_2} = \frac{\sum_{i=1}^{n_2} W_{2i} \uwd{x}_{2i}}{\sum_{i=1}^{n_2} W_{2i}}$. 
Note that $W_{1i},i=1,\ldots,n_1$ are unknown in models B, C and D, and for prediction, $N$ and $\uwd{\bar{X}}$ are unknown in all models.

Within the Bayesian
paradigm, $N$ and $\uwd{\bar{X}_p}$ are unknown parameters, and this uncertainty must also be accounted for.
Note that $W_{2i}, i=1,\ldots,n_2$, are known and are held fixed (not estimated) throughout the bootstrap procedure. Because the $W_{1i}$ are unknown weights and are estimated using the CLW procedure, they must be estimated at each step
of the bootstrap procedure. Now, it is our primary objective to incorporate the uncertainty in these estimated weights.
Specifically, we bootstrap the two samples $(\uwd{x}_{1i},y_{1i}),i=1,\ldots,n_1$ 
and $(W_{2i},\uwd{x}_{2i},y_{2i}),i=1,\ldots,n_2$, separately (i.e., for each $t$, we take a random sample of 
size $n_t$ with replacement). [Note again that $W_{1i},i=1,\ldots,n_1$, are unknown,
and must be calculated at each step of the bootstrap using the CLW procedure.]

The procedure has the following steps:
\begin{itemize}
\item[a.] Use the Bayesian bootstrap to draw a random sample from the nps and ps respectively. Note
          the Bayesian bootstrap is done with replacement.
					
\item[b.]	Use the method of CLW to estimate the  propensity scores.
					
\item[c.] Compute $\hat{N} =\sum_{j=1}^{n_2} W_{2j}$ 
          and $\uwd{\bar{x}_2} = \frac{\sum_{j=1}^{n_2} W_{2j} \uwd{x}_{2j}}{\sum_{j=1}^{n_2} W_{2j}}$.
\item[d.] Fit the models and do the prediction.		
\item[e.] Repeat (a), (b), (c) and (d) to get $B=1000$ bootstrap samples from the posterior distribution of the finite population mean.			
\end{itemize} 
Note that we are calculating the propensity scores in (b) using the method of CLW. This is how we account
for uncertainty in the estimation of the propensity scores.
	
\begin{table}[htb]
%\begin{scriptsize}
\begin{small}
\caption{Bootstrap study: Underestimation of variability for four selected models}
\label{Tab:boot}
\begin{center}
\begin{tabular}{ccccccccccccccccccccccccccc}
%\\
\hline
\\   
Model  &&   PM & PSD & CV  & NSE & 95\% CI\\
\\
\hline
\\
\multicolumn{2} {l} {\underline{No Bayesian bootstrap}}  \\
\\

    B && 27.321  & 0.153 &  0.002 &  0.006 & (27.029,  27.630)\\
		C && 27.045  & 0.134 &  0.002 &  0.005 & (26.787,  27.310)\\
	  D && 27.098  & 0.135 &  0.001 &  0.005 & (26.824,  27.350)\\
		E && 25.979  &  0.299 &  0.003 &  0.012 &  (25.395,  26.562)\\
    G && 26.856  &  0.304 &  0.011 &  0.009  &   (26.285,  27.470)\\
		
		\\
\multicolumn{2} {l} {\underline{Bayesian bootstrap}}  \\
\\
	
  B && 27.424  & 0.470 &  0.018 &  0.017 & (26.895,  27.845)\\
  C && 26.951  & 0.218 &  0.006 &  0.008 & (26.510,  27.334)\\
  D && 27.088  & 0.208 &  0.006 &  0.008 & (26.651,  27.436)\\
  E && 25.984  & 0.371 &  0.011 &  0.018 & (25.288,  26.772)\\
  G && 26.840  & 0.303 &  0.009 &  0.011 & (26.211,  27.374)\\

\\		
				\hline
\\

\end{tabular}
\flushleft
NOTE:  The bootstrap posterior distribution is based on 1000 samples that provide PM, posterior mean, PSD,
posterior standard deviation, $W$, width of the 95\% HPD interval and
CV, coefficient of variation. In C the prior is the nps and in D the prior
is the ps.
\end{center}
%\end{scriptsize}
\end{small}
\end{table}
\clearpage

To get the posterior density of $\bar{Y}$, letting $\uwd{\Omega}$ denote the vector of super-population parameters
and $[\cdot]$ denote distributions, we use the following decomposition for the joint distribution of all quantities,
$$
[\bar{Y}, \bar{X}, N, \uwd{\Omega},\uwd{W}_1, \uwd{W}_2, \uwd{y}_s] =
[\bar{Y} \mid  \bar{X}, N, \uwd{\Omega},\uwd{W}_1, \uwd{W}_2, \uwd{y}_s] \times
~[\uwd{y}_s \mid \uwd{W}_1, \uwd{W}_2, \uwd{\Omega}] \times
~[\uwd{\Omega}] \times ~[\bar{X}, N, \uwd{W}_1,  \uwd{W}_2],
$$
\begin{equation}
\label{boot}
[\bar{X}, N, \uwd{W}_1, \uwd{W}_2] = [\bar{X}, N\mid \uwd{W}_1, \uwd{W}_2] \times [\uwd{W}_1 \mid \uwd{W}_2] \times [\uwd{W}_2],
\end{equation}
where $[\bar{X}, N, \uwd{W}_1, \uwd{W}_2]$ is the bootstrap `posterior' distribution (i.e., where the
bootstrapping occurs).
 Equation (\ref{boot}) states  that once the bootstrap samples are obtained, simply fit all the models and do the predictions for every bootstrap sample. First, we draw samples from $[\uwd{W}_2]$ by bootstrapping 
$({W}_{2i},\uwd{x}_{2i},y_{2i}), i=1,\ldots,n_2$. Second, we draw samples from $[\uwd{W}_1 \mid \uwd{W}_2]$
by bootstrapping $(\uwd{x}_{1i}, \uwd{y}_{1i}), i=1,\ldots,n_1$, and running the CLW method to get 
$W_{1i},i=1,\ldots,n_1$. As in our general methodology, we condition on $\uwd{x}$.
%Throughout, we condition on $\uwd{x}$ and for the nps we assume independence of the sample indicators and the study variables, as in CLW.

In Table \ref{Tab:boot}, we compare Models B, C and D  to assess the underestimation in assuming the estimated weights
are known. The biggest increase is in B while C and D show less increases in standard deviations. For E,
the sample weights are known (not estimated) and there is still an  increase in standard deviation. Therefore, the
increase in standard deviation does not come only from assuming the estimated weights are known; yet this is more
significant.
%This phenomenon should be looked at more carefully.
Note $G$ does not have survey weights, except for $\hat{N}$ and $\bar{x}_2$ (weights are equal) from the ps.
	%
%\begin{table}[htb]
%%\begin{scriptsize}
%\begin{small}
%\caption{Bootstrap study: Underestimation of variability for four selected models}
%\label{Tab:boot}
%\begin{center}
%\begin{tabular}{ccccccccccccccccccccccccccc}
%%\\
%\hline
%\\   
%Model  &&   PM & PSD & CV  & NSE & 95\% CI\\
%\\
%\hline
%\\
%\multicolumn{2} {l} {\underline{No Bayesian bootstrap}}  \\
%\\
%
    %B && 27.321  & 0.153 &  0.002 &  0.006 & (27.029,  27.630)\\
		%C && 27.045  & 0.134 &  0.002 &  0.005 & (26.787,  27.310)\\
	  %D && 27.098  & 0.135 &  0.001 &  0.005 & (26.824,  27.350)\\
		%E && 25.979  &  0.299 &  0.003 &  0.012 &  (25.395,  26.562)\\
    %G && 26.856  &  0.304 &  0.011 &  0.009  &   (26.285,  27.470)\\
		%
		%\\
%\multicolumn{2} {l} {\underline{Bayesian bootstrap}}  \\
%\\
	%
  %B && 27.424  & 0.470 &  0.018 &  0.017 & (26.895,  27.845)\\
  %C && 26.951  & 0.218 &  0.006 &  0.008 & (26.510,  27.334)\\
  %D && 27.088  & 0.208 &  0.006 &  0.008 & (26.651,  27.436)\\
  %E && 25.984  & 0.371 &  0.011 &  0.018 & (25.288,  26.772)\\
  %G && 26.840  & 0.303 &  0.009 &  0.011 & (26.211,  27.374)\\
%
	%
%\\		
				%\hline
%\\
%
%\end{tabular}
%\flushleft
%NOTE:  The bootstrap posterior distribution is based on 1000 samples that provide PM, posterior mean, PSD,
%posterior standard deviation, $W$, width of the 95\% HPD interval and
%CV, coefficient of variation. In C the prior is the nps and in D the prior
%is the ps.
%\end{center}
%%\end{scriptsize}
%\end{small}
%\end{table}

It is desirable to do everything within the Bayesian paradigm. We should take care of the
variability of the unknown parameters in a single model, not in a two-stage procedure as
we have illustrated here using the additional bootstrap step. Unfortunately, this appears to
be a difficult computational problem.
%However, studying a 
%complete model within the Bayesian paradigm needs a different approach, much more
%difficult, and it is under study in a different research project, where we abandon the
%CLW method.

\vspace*{.25in}

\begin{center}
{\bf APPENDIX E: Importance of ps as Prior} 
\end{center}
\def\theequation{E.\arabic{equation}}
\setcounter{equation}{0}

We use a very simple model to help understand why when the ps is used to construct prior, the 
discounting is negligible. For historical data, consider
$$
y_{11},\ldots,y_{1n_1} \mid \mu, \sigma^2, a \stackrel{iid} \sim \mbox{Normal}(\mu, \frac{\sigma^2}{a}),
$$
$$
\pi(\mu,\sigma^2,a) \propto \frac{1}{\sigma^2}, -\infty<\mu<\infty, \sigma^2>0, 0<a<1. 
$$
Then, letting $\bar{y}_1$ and $s_1^2$ denote the sample mean and sample variance, and using Bayes' theorem,
\begin{equation}
\label{prior}
\pi(\mu,\sigma^2,a \mid \uwd{y}_1) \propto a^{n_1/2} (\frac{1}{\sigma^2})^{n_1/2+1} 
\exp\left\{-\frac{a}{2\sigma^2}\{(n_1-1)s_1^2 + n_1(\mu-\bar{y}_1)^2 \} \right\}.
\end{equation}
This is now the prior distribution of $(\mu,\sigma^2,a)$. When $a =1$, there is no
penalty, and this is what we mean by penalizing the ps or nps (historical data) (i.e.,
we are penalizing the prior distribution of $(\mu, \sigma^2)$ using historical data
(ps or nps)).

For the current data, consider
$$
y_{21},\ldots,y_{2n_2} \mid \mu, \sigma^2 \stackrel{iid} \sim \mbox{Normal}(\mu, \sigma^2).
$$
Then, using the historical prior in (\ref{prior}) and Bayes' therorem again, with
$\uwd{y} = (\uwd{y}_1,\uwd{y}_2)$,
$$
\pi(\mu,\sigma^2,a \mid \uwd{y}) \propto a^{n_1/2} (\frac{1}{\sigma^2})^{(n_1+n_2)/2+1} 
$$
\begin{equation}
\label{pos}
\times
\exp\left\{-\frac{1}{2\sigma^2}\{a(n_1-1)s_1^2 + an_1(\mu-\bar{y}_1)^2 
+ (n_2-1)s_2^2 + n_2(\mu-\bar{y}_2)^2 
\} \right\}.
\end{equation}

Let
$$
\hat{\mu} = \frac{an_1\bar{y}_1+n_2\bar{y}_2}{an_1+n_2}, 
\hat{\sigma}^2 = \frac{1}{an_1+n_2}. 
$$
Then, it is easy to show that
$$
\mu \mid \sigma^2,a,\uwd{y} \sim \mbox{Normal}(\hat{\mu}, \hat{\sigma}^2)
$$
and
$$
\sigma^2 \mid a, \uwd{y} \sim \mbox{InvGam}\left\{\frac{n_1+n_2-1}{2}, 
\frac{a(n_1-1)s_1^2 + an_1(\hat{\mu}-\bar{y}_1)^2 
+ (n_2-1)s_2^2 + n_2(\hat{\mu}-\bar{y}_2)^2}{2}\right\}.
$$

Finally, integrating $(\mu, \sigma^2)$ from (\ref{pos}), we have
$$
\pi(a \mid \uwd{y}) \propto \frac{a^{n_1/2}}{\sqrt{an_1+n_2}} 
$$
\begin{equation}
\label{a}
\times
\frac{1}{\left\{a(n_1-1)s_1^2 + an_1(\hat{\mu}-\bar{y}_1)^2 
+ (n_2-1)s_2^2 + n_2(\hat{\mu}-\bar{y}_2)^2\right\}^{(n_1+n_2-1)/2}}.
\end{equation}
Note that $\pi(a \mid \uwd{y})$ is well-defined for all values of $a$, $0\leq a \leq 1$.
%Now, assuming that $s_1^2 \approx s_2^2$, with common value $c^2$, and $|\hat{\mu}-\bar{y}_1| \approx |\hat{\mu}-\bar{y}_2|$, with common value $d^2$,
%(\ref{a}) becomes
%$$
%\pi(a \mid \uwd{y}) \approx \frac{a^{n_1/2}}{(an_1+n_2)^{(n_1+n_2)/2}} 
%\frac{1}{(c^2+d^2)^{(n_1+n_2)/2}}, 0 \leq a \leq 1.
%$$
If $n_1 < < n_2$, $\pi(a \mid \uwd{y}) \approx a^{n_1/2}, 0 \leq a \leq 1$, and 
$a$ will be close to $1$. Therefore, there will be very little discounting.
If $n_1 >> n_2$, $\pi(a \mid \uwd{y}) \approx \frac{1}{a^{n_2/2}}, 0 < a \leq 1$,
and $a$ will be close to $0$. Therefore, there will be significant discounting.
Approximately, only the relative magnitude of the sample sizes, $n_1$ and $n_2$, matters.
This explains why when the ps is used to construct the prior for the parameters, there
will be little discounting, and when the nps is used to construct the prior, there 
will be significant discounting.

%%%%%%%%%%%%%%%%%%%%%%%%%%%%%%%%%%%%%%%%%%%%%%%%%%%%%%%%%%%%%%%%%%%%%%%%%%%%%%%%%%%%%%%%%%%%%%%%%%
					
%%%%%%%%%%%%%%%%%%%%%%%%%%%%%%%%%%%%%%%%%%%%%%%%%%%%%%%%%%%%%%%%%%%%%%%%%%%%%%%%%%%%%%%%%%%%%%%%%%%%%%%%%

\end{document}